\documentclass[
superscriptaddress,
amsmath,amssymb,
aps, twocolumn, prb, floatfix
]{revtex4-2}

\usepackage[utf8]{inputenc}
\usepackage[T1]{fontenc}
\usepackage{mathptmx}
\usepackage{tabularx}
\usepackage[italian,english]{babel}
\usepackage[dvipsnames]{xcolor}
\usepackage[breaklinks,pdftex,pdfpagelabels,bookmarks]{hyperref}
\hypersetup{colorlinks=true,linktocpage=true,pdfstartpage=3,pdfstartview=FitH,breaklinks=true,pdfpagemode=UseNone,pageanchor=true,pdfpagemode=UseOutlines,plainpages=false,bookmarksnumbered, bookmarksopen=true,bookmarksopenlevel=1,hypertexnames=true,urlcolor=Brown,linkcolor=MidnightBlue!60!Black,citecolor=PineGreen}
\usepackage[version=4]{mhchem}
\usepackage{physics}
\usepackage{siunitx}
\usepackage{comment}
\usepackage{graphicx}
\usepackage{dcolumn}
\usepackage{bm}
\usepackage{amsmath}
\usepackage{mathtools, nccmath}
\usepackage{amsfonts}
\usepackage{amssymb}
\usepackage{xcolor}
\usepackage{bbold}

\renewcommand{\vec}{\bm}

\begin{document}
	
	\preprint{APS/123-QED}
	
	\title{Andreev non-Hermitian Hamiltonian for open Josephson junctions from Green's functions}
	
	\author{R. Capecelatro}
	\email{roberto.capecelatro@unina.it} 
	\thanks{These two authors equally contributed to this work.}
    \affiliation{Dipartimento di Fisica E. Pancini$,$ Università degli Studi di Napoli Federico II$,$ Monte S. Angelo$,$ via Cinthia$,$ I-80126 Napoli$,$ Italy}
	\author{M. Marciani}
	\email{marco.marciani89@gmail.com} 
	\thanks{These two authors equally contributed to this work.}
	\affiliation{Dipartimento di Fisica E. Pancini$,$ Università degli Studi di Napoli Federico II$,$ Monte S. Angelo$,$ via Cinthia$,$ I-80126 Napoli$,$ Italy}
	\author{G. Campagnano}
	\affiliation{CNR-SPIN$,$ UOS Napoli$,$ Monte S. Angelo$,$ via Cinthia$,$ I-80126 Napoli$,$ Italy}
	\author{P. Lucignano}
	\affiliation{Dipartimento di Fisica E. Pancini$,$ Università degli Studi di Napoli Federico II$,$ Monte S. Angelo$,$ via Cinthia$,$ I-80126 Napoli$,$ Italy}
	\begin{abstract}
        We investigate the transport properties of open Josephson junctions (JJs) through a minimal effective non-Hermitian (NH) approach derived from the equilibrium Green's function (GF) formalism. Specifically, we consider a JJ with a quantum dot barrier coupled to a normal metal reservoir. The coupling introduces an imaginary self-energy term in the JJ Hamiltonian which can be naturally accounted for in the NH formalism.
        While most approaches to similar problems work with the full junction Hamiltonian we propose a scheme for deriving an effective NH Hamiltonian for the Andreev levels only, that we compute from the singular part of the barrier GF. 
        To establish the range of applicability of this NH model we benchmark our results for both the dot density of states and the supercurrent against exact GF predictions in different transport regimes.
        We find that, as a rule of thumb, the Andreev NH description is accurate when the spectral overlap between the Andreev bound states (ABS) and the near-gap continuum states is negligible, i.e. when the ABS energies lie sufficiently far from the superconducting gap relative to their line-width.
        This method not only highlights the effective physics of the JJ but also offers a scalable framework for studying large-size devices.
	\end{abstract}

	\maketitle
	
	\section{Introduction}

Quantum mechanics is governed by Hermitian operators, ensuring real eigenvalues and unitary time evolution that preserve probability.
However, theoretical and experimental developments have generated interest in non-Hermitian (NH) quantum systems~\cite{Gong_2020,Oku23}, characterized, in general, by complex eigenvalues and non-orthogonal eigenmodes, challenging the traditional paradigm of quantum mechanics~\cite{Bender07,Bender:1998}.

NH Hamiltonians provide a natural framework for describing open systems that exchange energy with their environment. The imaginary part of the eigenvalues embodies gains and losses, whereas non-orthogonality of the eigenvectors unlocks interesting phenomena such as the energy exchange among different modes near the system degeneracies~\cite{Doppler2016,Xu2016}, the NH-skin effect~\cite{FoaTorres2018, Yao18, Zhang:2022}, where modes concentrate near the system boundary in response to asymmetric interactions, and novel topological properties~\cite{Kun18,Hui18,Kaz19,Zhe20,Budich2020, FoaTorres2020, Nakamura2023, Manna2023}. 

At present, NH physics has been investigated in a number of different platforms ranging from optics~\cite{Pen14, Doppler2016, St-Jean2017,Zhou2018,El-Ganainy2019} and plasmonics ~\cite{alaeian2014, lourencco2018} to molecular mechanics~\cite{Lef09} and  optomechanics~\cite{schonleber2016,Xu2016}, 
where one can access a precise control over few spectrally-isolated modes and their decay channels, thus finely tuning gains and losses. 

Recent advancements have also extended NH physics to solid-state systems~\cite{Zhang:2021}. In particular, there has been a substantial interest in exploring the impact of NH physics in different condensed matter settings such as 
mesoscopic hetero-structures involving topological insulators \cite{Philip2018, Chen2018, Bergholtz2019, Kawabata2019, Budich2020, Cayao2023_2}, topological~\cite{Mi14,SanJose2016, Avila:2019, Cayao2023, Arouca2023, Sayyad2023, Cayao2024, CayaoAguado2024} and parity-time (PT) symmetric superconductors~\cite{Kawabata2018, Cayao2022, Kornich2022, Kornich2022_2, Kornich2023}. 

Among these systems, Josephson junctions (JJs) and superconducting quantum devices offer the possibility of a precise control over quantum coherence of isolated degrees of freedom~\cite{Josephson1962, Barone1982, DeGennes, Tinkham, Devoret1984, Martinis1987, Clarke1988, Cleland1988, Devoret2013, Flensberg2021, Harrington2022}, enabling experiments that explore complex quantum phenomena in controlled environments~\cite{Avila:2019}.

In particular, single- or multi-terminal JJs with tunable couplings with external metallic electronic reservoirs reveal to be intriguing platforms to explore in depth the interplay of superconductivity and NH physics \cite{Melin2022, Kornich2022, Kornich2023, Shen2024, 
Cayao2023, Cayao2024, CayaoAguado2024, CayaoSato2024, CayaoSato2024_2, Pino2024, Li2024, Beenakker2024}. Non-Hermitian behavior in JJ systems has been achieved by introducing effective gain and loss via engineered environments, yielding insights into dissipation-driven phase transitions and quantum state localization~\cite{Naghiloo:2019}.

Open Josephson junctions, with normal leads, have been extensively studied with Green's functions techniques, scattering matrix and master equation approaches~\cite{Buttiker1985, Wees1991, Itoh1995, Chang1997, Brouwer1997, Samuelsson1997, Belogovskii2001, Samuelsson2001, Lantz2002, Schapers2003, JonckhereeMartin2009, Weiss, BreuerPetruccione, Caldeira1983b, Caldeira2014}. 
In equilibrium conditions no net
supercurrent can flow on the
normal leads but they still affect the JJs by inducing decoherence, lowering the intensity of the Josephson current~\cite{Buttiker1985, Wees1991, Itoh1995, Chang1997, Brouwer1997, Samuelsson1997, Belogovskii2001, Samuelsson2001, Lantz2002, Schapers2003, JonckhereeMartin2009}.

A NH description of these setups is useful as it can provide us with some interesting insights on the effective physics governing them. In this context, a NH Hamiltonian for the JJ has been derived both with Green's function (GF)~\cite{Melin2022, Shen2024, Pino2024, Cayao2023, Cayao2024, CayaoSato2024, CayaoSato2024_2} and scattering matrix (SM) methods~\cite{Li2024, Beenakker2024}. One of the focal points of several recent theoretical studies has been the connection between the complex eigenvalues of the NH Hamiltonian and the supercurrent, along with other thermodynamic observables.

Adopting a GF approach, the influence of a normal lead on the JJ is accounted for in an imaginary self-energy term that is often assumed frequency independent under the so-called "broad-band" approximation, the junction NH Hamiltonian reading as $H_{eff}^{NH}=H_{JJ}+\Sigma_{N}(\omega=0)$.

The authors in Ref.~\cite{Shen2024} demonstrated how to compute from a NH Hamiltonian of this form both the Josephson current, i.e. $J(\phi)\propto \partial_{\phi} \mathrm{Im}\,\mathrm{Tr}\,H_{eff}\mathrm{ln}H_{eff}$ at $T=0$, and the current susceptibility.
Other works propose to separate the contribution of the real and imaginary parts of the NH Hamiltonian eigenvalues to the Josephson current, with the former describing the physical current flowing through the JJ and the latter accounting for the losses, thus simply extending the well known formula of the closed junction limit, i.e. $J(\phi) \propto \sum_{n}\partial_{\phi}(\mathrm{Re}E_{n}(\phi) -i \partial_{\phi}\mathrm{Im}E_{n}(\phi))$~\cite{Cayao2023, Cayao2024, CayaoSato2024}.
Both these approaches are non-perturbative and are valid so long as the broad-band limit holds. However, they apply to NH Hamiltonians in tight-binding form which comprise both the barrier and the S leads, hence being of difficult implementation for large size or multi-channels JJs ~\cite{Cayao2023, Cayao2024, CayaoSato2024, Shen2024}.
 
In this regard, we note that the transport properties of multi-terminal JJ setups can be described in terms of the barrier GF only (or the barrier SM) that includes the influence of both the S and N leads as, in general, frequency-dependent self-energy terms. This allows an agile handling of large size JJs by tracing out all leads degrees of freedom. 
Moreover, in the closed JJ limit, i.e. without N leads, in specific regimes, e.g. the short-junction one, the Josephson current is mainly due to the sub-gap Andreev bound states (ABS), with energies $\epsilon_A(\phi)$ corresponding to poles of the barrier GF (or SM), with a negligible supra-gap continuum contribution~\cite{Beenakker1992, Beenakker1991_A, Beenakker1991_B, Furusaki_Takayanagi_Tsukada_1992, Furusaki_Tsukada_1991, Sellier2005, Benjamin2007, Meng2009_PRB, JonckhereeMartin2009, Capecelatro2023}, i.e. $J\propto \sum_{\varepsilon_{A}} \partial_\phi \varepsilon_A(\phi)$.

It is then natural to ask if such an ABS description can be valid also in the open junction case where, due to the coupling with the N leads, the barrier GF (or SM) poles acquire an imaginary part that manifests as a broadening of the Andreev levels~\cite{Pino2024, Li2024, Beenakker2024}. Further, it is worth asking if a NH Hamiltonian for the barrier only can be derived from the barrier GF (or SM) and it is sufficient to describe the JJ transport properties.
Several current formulas involving such barrier GF (or SM) poles have been derived for slightly different systems~\cite{Pino2024, Cayao2024, Li2024, Beenakker2024}, often referring to specific coupling regimes, and a comprehensive study of their applicability is still missing.

In this manuscript, we aim at identifying a NH Hamiltonian for the Andreev levels only which suffice to describe the JJ spectral and transport properties.
We consider a simple system consisting of a JJ with a single-level quantum dot (QD) barrier coupled to a normal electron reservoir (SQDNS JJ). A sketch of the device is shown in Fig.~\ref{fig: 1_SQDNS}. 
We investigate whether the dot GF poles identified as the complex energies of the Andreev \emph{quasi-bound} states (quasi-ABS) are responsible for the supercurrent. Further, we propose a scheme for building the effective \emph{Andreev} NH Hamiltonian of the dot from the singular part of its GF. This allows the derivation of a generalized NH current formula for the quasi-ABS contribution. Contextually, we also compute the current by exploiting our Andreev NH Hamiltonian in the current formula of Ref.\cite{Shen2024}, similarly to what is done in Ref.\cite{Beenakker2024} for a weakly-coupled quantum dot.
Importantly, we refer to results obtained in the exact Green’s functions formalism as a benchmark to discuss the validity of the \emph{effective} NH Hamiltonian approach in different transport regimes.

We show that our generalized current formula describes accurately the sub-gap current provided that the sub-gap contribution from the continuum, that is produced near the gap by the hybridization between the dot and the N lead, is negligible. Mathematically, this happens when the full dot GF can be approximated by its singular part. 
In practice, such effective NH approach is found to succeed when there is no overlap in the spectral density between the quasi-ABS and the continuum quasiparticles. This in turn depends on whether the line-width of the sub-gap ABS is small or large compared to their distance from the superconducting gap. 
To conclude, we find that the effective Andreev NH Hamiltonian exactly describes the system observables as long as the quasi-ABS energies are far from the gap. 

This work is organized as follows. In Section~\ref{sec_2_Model} we present the Hamiltonian of the junction and the GF of the quantum dot. In Section~\ref{sec_3_Transport_Properties} we recall how to compute the Josephson current in the Green's functions formalism analyzing the effect of the normal reservoir on the current-phase relation (CPR) by varying its coupling with the dot in two relevant regimes corresponding to the situations in which the dot is weakly and strongly coupled to the S leads. 
We study the analytical properties of the dot GF in Sections~\ref{sec: green_decomp} and \ref{sec: polarAndNH}, where we also show how to build the Andreev NH Hamiltonian from the dot GF. In Section~\ref{sec: dos}, we analyze the poles of the dot GF in both weak- and strong-coupling regimes. In that context, we show how the broadening of the ABS energies due to the coupling with the reservoir affects the QD density of states (DOS).
In Section~\ref{sec: Jpol and JHeff formulas} we derive our current formula and consider an adaptation of the current equation of Ref.\cite{Shen2024} to our case. Finally, we compare these two effective current models with the exact results from the Green's functions technique in Section~\ref{sec: Jpol and JHeff numerics}.

\section{Model}\label{sec_2_Model}
\subsection{Hamiltonian}
\begin{figure}[th]
		\centering
		\includegraphics[scale=0.15]{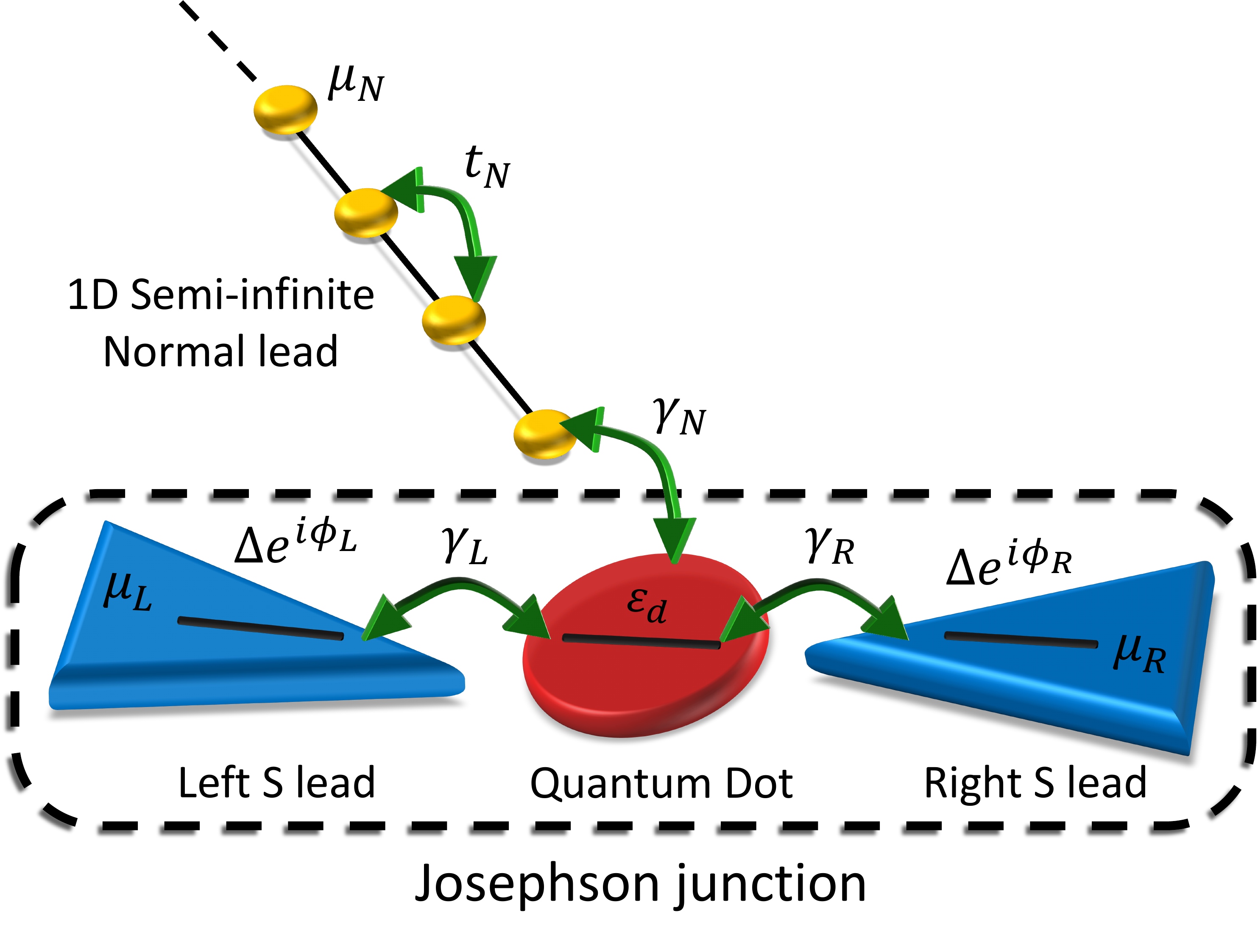}
		\caption[Scheme of the T-shaped junction]{Scheme of the SQDNS JJ consisting of two bulk s-wave superconducting leads, a single-level quantum dot barrier and a 1D semi-infinite normal lead. $\varepsilon_{d}$ is the energy of the quantum dot level. $t_N$ is the hopping amplitude along the 1D normal chain. The chemical potential of each lead and the hopping amplitudes between the S and N leads and the dot are $\mu_{L}$, $\mu_{R}$ and $\mu_{N}$, and $\gamma_{L}$, $\gamma_{R}$ and $\gamma_{N}$, respectively. The two superconductors have equal superconducting gap $\Delta$ and differ only by their superconducting phases, $\phi_{L/R}$.}
	\label{fig: 1_SQDNS}
\end{figure}
The system under study consists of a T-shaped superconductor - quantum dot - superconductor (SQDS) Josephson junction (JJ) with a single level QD that is connected to a 1D semi-infinite normal lead (SQDNS JJ), as in Fig.~\ref{fig: 1_SQDNS}, acting as a fermionic reservoir.

The Hamiltonian  reads 

\begin{equation}
		\label{Hamiltonian_system}
		H_{\rm sys}=H_{\rm D}+ H_{\rm S } + H_{\rm N }+ H_{\rm T}\,,
\end{equation}

where $H_{\rm D}$, $H_{\rm S }$, $H_{\rm N}$, are the dot, the S leads, and the N lead Hamiltonians, respectively, and $H_{\rm T}$ is the tunneling Hamiltonian between the leads and the dot.
	
The dot Hamiltonian is 
\begin{equation}
		H_{\rm D}=\varepsilon_{d}\sum_{\sigma=\uparrow,\downarrow}d_{\sigma}^{\dagger}d_{\sigma},
		\label{H_Quantum_Dot}
\end{equation}
where $\varepsilon_{d}$ is the dot energy that can be controlled by an external gate voltage, $d_{\sigma}$ denotes the annihilation operator for electrons of spin $\sigma=\uparrow,\downarrow$ on the dot.

Following Refs.\cite{Benjamin2007, Meng2009_PRB, JonckhereeMartin2009, Capecelatro2023, Beenakker1992, Beenakker2024}, we neglect Coulomb repulsion on the dot, which is usually disregarded in the limit of large coupling between the leads and the dot~\cite{Beenakker1992, Glazman1989,Benjamin2007, Meng2009_PRB, JonckhereeMartin2009, Capecelatro2023}.
The superconducting electrodes are s-wave and have equal chemical potential $\mu$, normal-state dispersion $\varepsilon_{\vec{k},\sigma}$ and superconducting gap $\Delta$. The S leads Hamiltonian then reads:
\begin{equation}
		\begin{split}
			H_{\rm S \,}=&	\sum_{i=L,R} H_{\rm i}=\sum_{i=L,R}\sum_{\vec{k}}\sum_{\sigma=\uparrow,\downarrow}(\varepsilon_{\vec{k},\sigma}-\mu) c_{i,\vec{k},\sigma}^{\dagger}c_{i,\vec{k},\sigma}+ \\
			&+\sum_{i=L,R}\sum_{\vec{k}} \Delta e^{i\phi_{i}}c_{i,\vec{k},\uparrow}^{\dagger}c_{i,-\vec{k},\downarrow}^{\dagger} +\rm H.c. \hspace{0.3 mm},
		\end{split}
		\label{H_leads}
\end{equation}

where $c_{i,\vec{k},\sigma}$ represents the annihilation operator for electrons in the state $\vec{k}$ with spin $\sigma$ on the lead $i$ ($i=L,R$).
Here, $\phi_{i}$ is the superconducting phase in the lead $i$. 

The normal lead consists in a 1D semi-infinite chain whose tight-binding Hamiltonian $H_{\rm N}$ reads:
\begin{equation}
        H_{\rm N}=\sum_{\vec{k}}\sum_{\sigma=\uparrow,\downarrow}(\mu_N-\varepsilon_{N,\,\vec{k},\sigma}) c_{N,\vec{k},\sigma}^{\dagger}c_{N,\vec{k},\sigma}\,,
        \label{H_N_lead}
\end{equation}
with $\mu_{N}$ being the normal metal chemical potential and $\varepsilon_{N,\, \vec{k},\sigma}=-2 t_{N} \cos{\left(k a_{N}\right)}$ being the dispersion law in momentum space. $t_{N}$ and $a_{N}$ are the hopping parameter and lattice constant of the 1D chain, respectively.

The tunneling Hamiltonian $H_{\rm T}$ includes the interaction between the dot and both the superconductors and the N lead, thus reading
 \begin{equation}
		H_{\rm T}= \sum_{i=L,R, N} H_{\rm T_{i}}=\sum_{i=L,R, N} \gamma_{i} \sum_{\vec{k}}\sum_{\sigma}  c_{i,\vec{k},\sigma}^{\dagger}d_{\sigma} +\rm H.c. \hspace{0.3 mm},
		\label{H_hopping}
\end{equation}
where $\gamma_{i=L,R,N}$ are the hopping amplitudes, see Fig.~\ref{fig: 1_SQDNS}, assumed to be real and $\vec{k}$-independent.
Since we are interested interested in the equilibrium properties of this Josephson system, we set $\mu = \mu_N =0$ in the following.
	
\subsection{Quantum dot Green's function}

In the diagrammatic approach, the \emph{dressed} Green’s function of the dot, embodies the coupling with the S electrodes and the N reservoir. It enters the calculation of both the CPR and DOS of the JJ thus encoding all its spectral and transport properties at equilibrium.
To compute the dot Matsubara GF, $G_{d}(i\omega_{n})$, we write a closed Dyson equation including its \emph{bare} GF, $G_{d}^{0}(i\omega_{n})=(i\omega_{n}-H_{\rm D})^{-1}$, and its interactions with both the leads in a self-energy term, i.e. 
$\Sigma(i\omega_{n})=\sum_{i=L,R,N} \Sigma_{i}(i\omega_{n})=\sum_{i} H_{\rm T_{i}}G^{0}_{i}(i\omega_{n})H_{\rm T_{i}}^{T}$, where $G^{0}_{i}(i\omega_{n})=(i\omega_{n}-H_{\rm i})^{-1}$ is the \emph{bare} GFs of the lead $i=L,R,N$. The Dyson equation for $G_{d}(i\omega_{n})$ in Nambu representation reads

\begin{equation}
		\hat{G}_{d}(i\omega_{n}) = \left(i\omega_{n}\hat{1}-\hat{H}_{D}-\sum_{i=L,R,N} 	\hat{H}_{T_{i}}\hat{G}^{0}_{i}(i\omega_{n})\hat{H}_{T_{i}}^{T}\right)^{-1} \,,
		\label{Gd_Dyson_equation_explicit}
\end{equation}
where $\omega_{n}=\pi(2n+1)T$ with $T$ being the temperature and we adopted units with $\hbar=k_B=1$, where $\hbar$ is the reduced Planck constant, and $k_B$ is the Boltzmann constant. Hereafter, we use the hat symbol $\hat{\cdot}$ to indicate $2\times2$ matrices in the Nambu-space.

The GF of the S leads  can be expressed  in a simplified form  assuming that they are described by a flat conduction band with a constant density of state $\rho_{0}$ at the Fermi level~\cite{Cuevas1996, Cuevas2001, JonckhereeMartin2009, Capecelatro2023, Glazman1989, Sellier2005, Benjamin2007, Meng2009_PRB}. Under this assumption, following Refs.~\cite{Cuevas1996, Cuevas2001, JonckhereeMartin2009, Capecelatro2023, Glazman1989, Sellier2005, Benjamin2007, Meng2009_PRB}, $\hat{G}^{0}_{i=L/R}(i\omega_{n})$ reads 
\begin{equation}
		\label{Bare_lead_GF_chap_4}
		\begin{split}
			\hat{G}^{0}_{i}(i\omega_{n})=&\frac{-i\omega_{n}\pi \rho_{0}}{\sqrt{\Delta^2-(i\omega_{n})^2}}
			\hat{1} +\frac{i\Delta\pi \rho_{0}}{\sqrt{\Delta^2-(i\omega_{n})^2}}\times\\ &\times\left[e^{i\phi_{i}} \left(\frac{\hat{\tau}_{x}+i\hat{\tau}_{y}}{2}\right) - e^{-i\phi_{i}} \left(\frac{\hat{\tau}_{x}-i\hat{\tau}_{y}}{2}\right)\right]\, ,
		\end{split}
\end{equation}
with $\hat{\tau}_{j=x,y,z}$ being the equivalent of Pauli matrices in Nambu space. 

The GF of a normal semi-infinite lead can be computed in tight-binding approach~\cite{Ferry2009:book,JonckhereeMartin2009, Datta_1995, Ryndyk2009}, see Appendix~\ref{app: GFs_of_the_leads}, and is written as
	\begin{equation}
		\label{Bare_N_lead_GF}
		\hat{G}^{0}_{N}(i\omega_{n}, \mu_{N}=0)=\frac{i\omega_{n}}{2t_{N}^2}\hat{1}-\frac{i \mathrm{sign}(\omega_{n})}{t_{N}^2}\sqrt{t_{N}^2-\frac{\left(i\omega_{n}\right)^2}{4}}\hat{1}\, .
	\end{equation}
 
We can derive an explicit and compact expression for the dot GF by introducing $\Gamma=2\pi\rho_{0}\gamma^2$ and $\Gamma_{N}=\gamma_{N}^{2}/t_{N}$, describing the dot-superconductors and the dot-normal metal hybridizations, respectively
\begin{eqnarray}
		\label{Gmatsu}
		\hat{G}_{d}(i\omega_{n})& = & \Big[ i\omega_{n}\left(1+\frac{\Gamma}{\sqrt{\Delta^2-(i\omega_{n})^2}}\right)\hat{1} +\varepsilon_{d}\hat{\tau}_{z}+\\ 
	&&+\frac{\Gamma\Delta}{\sqrt{\Delta^2-(i\omega_{n})^2}} \cos{\left(\frac{\phi}{2}\right)}\hat{\tau}_{x}-\nonumber\\
		&&-\Gamma_{N}\left(\frac{i\omega_{n}}{2 t_{N}}-i  \sqrt{1-\frac{\left(i\omega_{n}\right)^2}{4t_{N}^2}}\mathrm{sign}(\omega_{n})\right)\hat{1}\Big]^{-1} \,,
		\nonumber
\end{eqnarray}
where we set $\phi_L = -\phi_R = \phi/2$. 

The expression for $\hat{G}_{N}^{0}$ gets simplified under the so-called \emph{broad-band} approximation, by which the metal produces no retardation. This is valid when $t_{N}\gg\omega_{n}$, so that we can take $i\omega_{n} \rightarrow 0$ in Eq.~\ref{Bare_N_lead_GF}. The associated self-energy term in the dot GF reduces to an imaginary term $	\Sigma_{N}=H_{\mathrm{T_{N}}}G^{0}_{N}H_{\mathrm{T_{N}}}^{T}=-i(\gamma_{N}^{2}/t_{N})\mathrm{sign}\left(\omega_{n}\right)=-i\Gamma_{N}\mathrm{sign}\left(\omega_{n}\right)$. 
Intuitively, this term affects the junction only by inducing a broadening of the dot energy proportional to $\Gamma_{N}$.

In a closed junction, with no N lead, the GF poles are the real energies of the sub-gap Andreev Bound States (ABS) and from their knowledge the sub-gap supercurrent is directly computed~\cite{Beenakker1991_A, Beenakker1991_B, Furusaki_Tsukada_1991, Furusaki_Takayanagi_Tsukada_1992, Benjamin2007, Meng2009_PRB, Krichevsky_PRB_2000, Beenakker1992, Capecelatro2023}. When a normal metal reservoir is attached to the dot, the $\hat{G}_{d}$ poles acquire an imaginary part as a consequence of the dot level broadening~\cite{CayaoAguado2024, Cayao2023, Cayao2024, CayaoSato2024, Li2024, Shen2024, Pino2024, Beenakker2024}.
In the following sections, we investigate the analytic structure of the dot GF in the open system case. We analyze in which way and to what extent the complex poles can be exploited to predict the spectral and transport properties of the junction, Section~\ref{sec: poles analytic expression}.
Hereafter, we address these complex poles as Andreev \emph{quasi-bound} states (quasi-ABS) with their real part describing their energy and the imaginary part their line-width.

We note that the numerics will be performed using the full frequency dependence in the N lead GF, to highlight the generality of the method. However, we choose a regime of parameters, i.e. $t_{N}\gg \omega_{n}, \Delta, \Gamma, \Gamma_{N}$, at which the broad-band approximation is valid and does not affect the transport, so as to enable a comparison with analytical expressions in Sec.~\ref{sec: poles analytic expression}.

 \section{Transport properties with Green's function techniques} \label{sec_3_Transport_Properties}
 \subsection{Current formula}
	\label{sec: CurrForm}
     
In closed systems the supercurrent is tightly related to the GF poles~\cite{Beenakker1992}, this connection in open junctions is still the object of current investigations~\cite{Li2024, Cayao2023, Cayao2024, CayaoSato2024, CayaoAguado2024, Pino2024, ohnmacht2024}. The current flowing through multi-terminal junctions with normal and superconducting leads, can be either derived by means of the Green's function or scattering matrix formalisms~\cite{JonckhereeMartin2009, Itoh1995, Samuelsson1997, Samuelsson2001, Lantz2002, Schapers2003}. In the absence of a bias voltage between the leads no current can flow from the dot toward the normal lead and the normal current between the S leads identically vanishes, see Appendix~\ref{Current in the SQDNS JJ}. Therefore, the supercurrent driven by the phase difference between the two S leads  is~\cite{Asano2019, JonckhereeMartin2009, Furusaki1994, Asano2001, Minutillo2021, Ando_1991, Ahmad2022, Cuevas1996, Cuevas2001, Clerk2000, Benjamin2007, Meng2009_PRB,  Capecelatro2023}
   \begin{equation}
		\label{Josephson_Current_Matsubara_text}
		J(\phi)= \dfrac{i e T}{2}  \sum_{\omega_{n}} \mathrm{Tr} \left[\hat{\tau}_{z}\left(\hat{H}_{T_{R}}\hat{G}_{d}\hat{H}_{T_{R}}\hat{G}_{R}^{0} -\hat{H}_{T_{R}}\hat{G}_{R}^{0}\hat{H}_{T_{R}}\hat{G}_{d} \right) \right] \, ,
	\end{equation}
    where $\mathrm{Tr}$ stands for the trace over the Nambu space. Eq.~\ref{Josephson_Current_Matsubara_text}, when making explicit the S lead GF in Eq.~\ref{Bare_lead_GF_chap_4}, simplifies to

\begin{equation}
		\label{Current_Matsubara_Final_chap_4}
		J\left(\phi\right)=  e T\Delta \Gamma \sin\left(\frac{\phi}{2}\right) \sum_{\omega_{n}}\frac{\Re{F_{d}(\omega_{n})}}{\sqrt{\Delta^2+\omega_{n}^2}} \; ,
\end{equation}
	
where $ F_{d}=\left(\hat{G}_{d}\right)_{12}$ is the dot \emph{anomalous} GF, describing the superconducting correlations induced on the dot by proximity effect, see Appendix~\ref{Current in the SQDNS JJ} for details.
    
In closed systems, it is sufficient to integrate separately the sub-gap and supra-gap currents densities to compute, respectively, the ABS and the continuum quasiparticles contributions.
In the zero temperature limit, when substituting $i\omega_{n}\rightarrow \omega+ i\eta$ and taking the limit $\eta\rightarrow0^{+}$, these can be expressed~\cite{Benjamin2007,Capecelatro2023,Sellier2005, Clerk2000, Meng2009_PRB}  in terms of the real and imaginary parts of $F_{d}^{R}\left(\omega\right)$, respectively:
	\begin{equation}
		\label{Andreev levels current}
		J^{in}=J_{|\omega|<\Delta}=2e\Gamma\Delta\sin{\left(\frac{\phi}{2}\right)}\int_{-\Delta}^{0}\frac{d\omega}{2\pi}\frac{\Im{F_{d}^{R}\left(\omega\right)}}{\sqrt{\Delta^2-\omega^2}}\, ,
	\end{equation}
	\begin{equation}
		\label{Quasiparticles current}
		J^{out}=J_{|\omega|>\Delta}=2e\Gamma\Delta\sin{\left(\frac{\phi}{2}\right)}\int_{-\infty}^{-\Delta}\frac{d\omega}{2\pi}\frac{\Re{F_{d}^{R}\left(\omega\right)}}{\sqrt{\omega^2-\Delta^2}}\, .
	\end{equation}
This correspondence does not automatically hold for an open junction.
When the dot is connected to the normal lead, the continuum of N, with a finite density of states inside the gap, hybridizes with both the Andreev levels and the S leads continuum. Such mechanism generates a sub-gap continuum in which the two contributions are mixed. In this case, these formulas only measure the sub-gap and supra-gap currents, that we here refer to as $J^{in}$ and $J^{out}$, respectively.

In Secs. \ref{sec: green_decomp}, \ref{sec: poles analytic expression} and \ref{sec: JPolar and JHeff} we investigate how the GF poles, related to the ABS, contribute to the sub-gap current.
 
\subsection{Weak and strong-coupling regimes}
    \begin{figure}
	   \centering \includegraphics[scale=0.38]{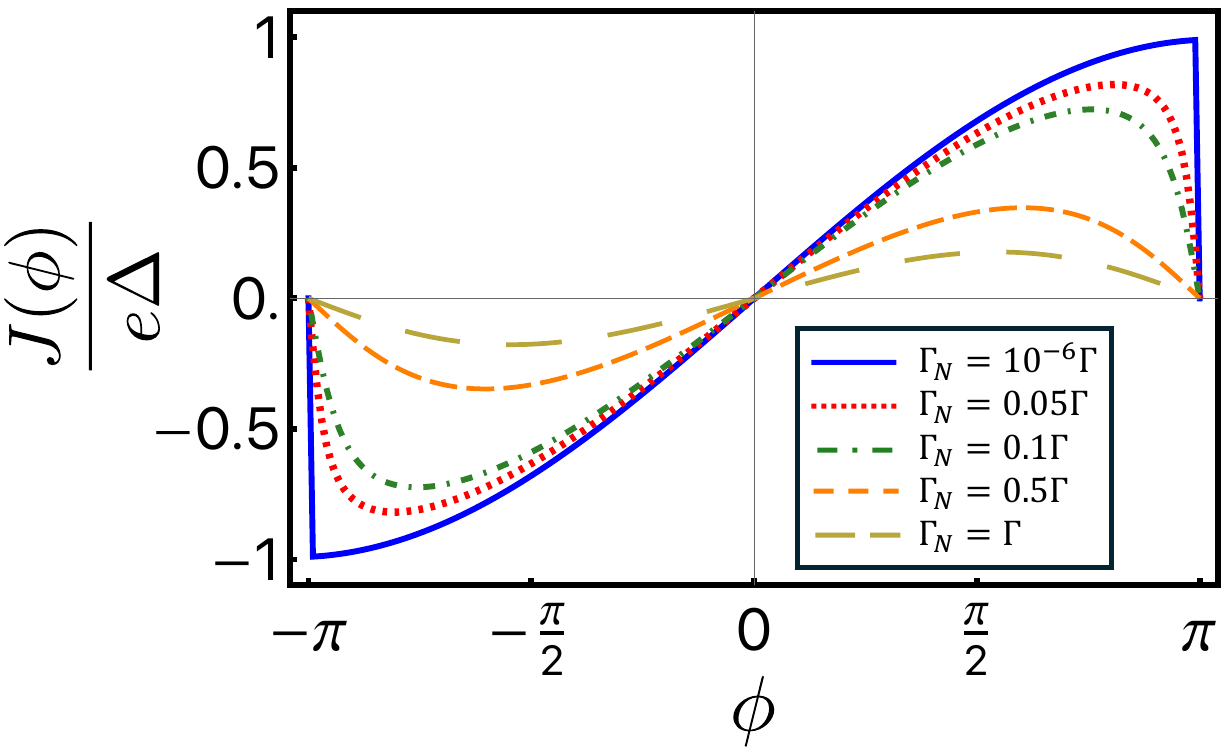}
	   \caption{Current-phase relation (CPR) of the junction in the strong-coupling regime by varying the amplitude of the dot - N lead hybridization parameter $\Gamma_{N}$. The system parameters are (in units of $\Gamma$) $\varepsilon_{d}=0$, $\Delta=0.01$, $t_N=10$ and $T=10^{-3}\Delta$.}
	   \label{fig: 2_Current_Matsubara_SC}
    \end{figure}
 
    \begin{figure}
        \centering
        \includegraphics[scale=0.56]{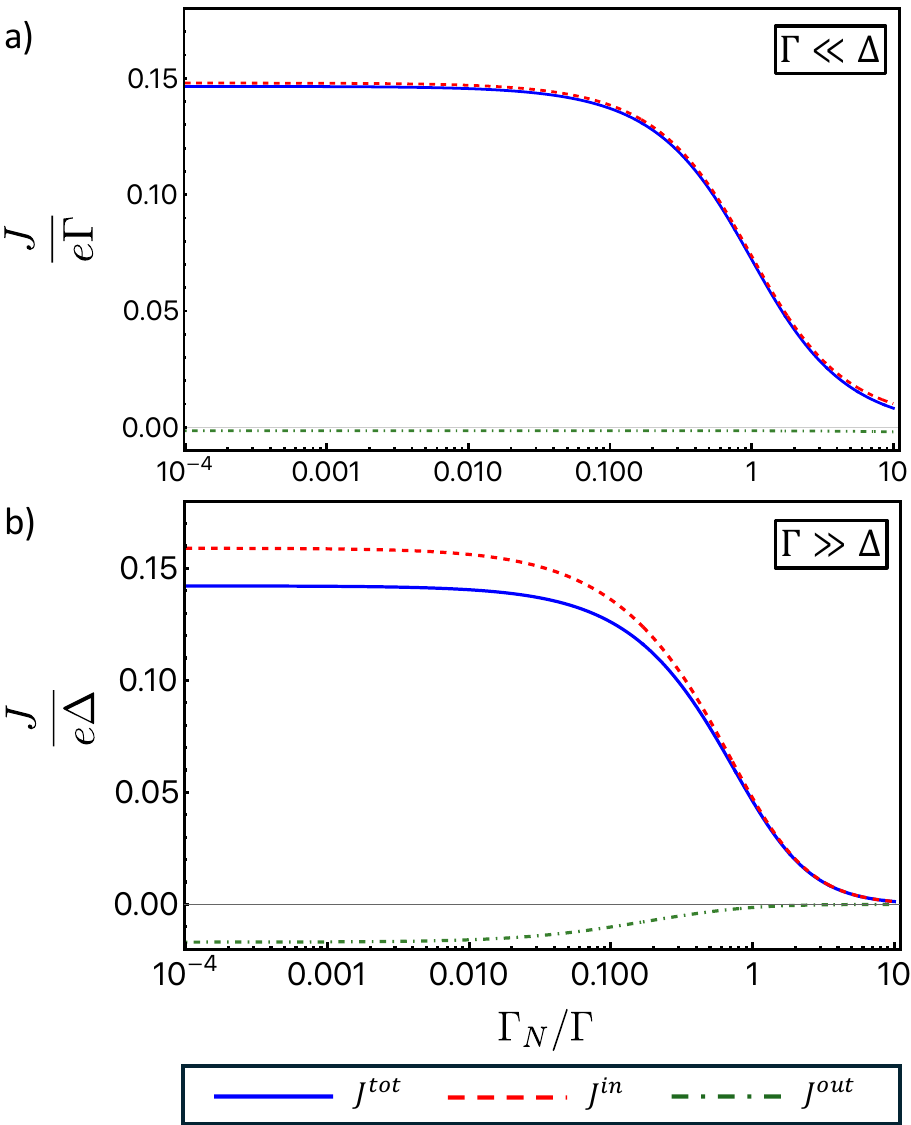}
        \caption{Sub-gap and supra-gap current contributions as a function of the ratio $\Gamma_{N}/\Gamma$ at fixed phase bias $\phi=0.3 \,\mathrm{rad}$ for the weak (a) and strong (b) coupling regimes. In the insets the ratio between $J_{out}$ and $J_{in}$. The other parameters are $\varepsilon_{d}=0$ and for the weak-coupling case (in units of $\Delta$) $\Gamma=0.01$, $t_N=10$ and $T= 10^{-3}$ while for the strong-coupling one (in units of $\Gamma$) $\Delta=0.01$, $t_N=10$ and $T= 10^{-3}$.}
        \label{fig: 3_Jin_Jout_panel}
    \end{figure}

In this section, we analyze the effect of the N lead on the Josephson current in both
situations in which the hybridization of the dot with the S leads is much larger or smaller than the superconducting gap, i.e. $\Gamma\gg\Delta$ or $\Gamma\ll\Delta$~\cite{Beenakker1992, Capecelatro2023}, also known in literature as narrow- and wide-resonance limits~\cite{Beenakker1991_B, Beenakker1992}. We define these limits as weak- and strong-coupling regimes, respectively.

In the closed junction limit, they are characterized by a very low continuum QP contribution to the current, thus their CPR behavior is mainly ascribed to the sub-gap ABS current. 

In the former case, the ABS are formed at energy scales of $\Gamma\ll\Delta$ while in the latter they occur at energies $\sim\Delta$~\cite{Beenakker1992, Golubov2004, Benjamin2007, Capecelatro2023, Furusaki_Takayanagi_Tsukada_1992}.
These transport regimes present two different physical situations in which the bound states are very far and very close to the superconducting gap, respectively. They provide us with a simple and interesting playground to test whether quasi-ABS description is valid for the open junction case.

We notice that the intermediate coupling regime, at $\Delta\sim\Gamma$, is characterized by a sizable supra-gap continuum contribution to the Josephson current even in the closed JJ case. For this reason, this regime is not suitable for being described by an effective NH model based on the Andreev levels only and, thus, it is not analyzed in this work.
For the sake of simplicity, hereafter we focus our attention on the resonant tunneling regime, thus we set $\varepsilon_{d}$ is equal to the leads chemical potential. We note that under this condition the hybridization parameter $\Gamma$ with the superconductors plays the role of the Thouless energy~\cite{Beenakker1991_B, Beenakker1992}. Therefore, at $\varepsilon_{d}=\mu$ the strong-coupling regime, i.e. $\Gamma\gg\Delta$, corresponds to the short-junction limit~\cite{Beenakker1991_B, Beenakker1992, Giuliano2013, Giuliano2014}, which is of particular interest for circuital applications. 
We numerically analyze the CPR behavior of the JJ also at $\varepsilon_{d}\neq0$ in Appendix~\ref{app: epsd neq}.

    \begin{figure}[hbtp]
        \centering
        \includegraphics[scale=0.68]{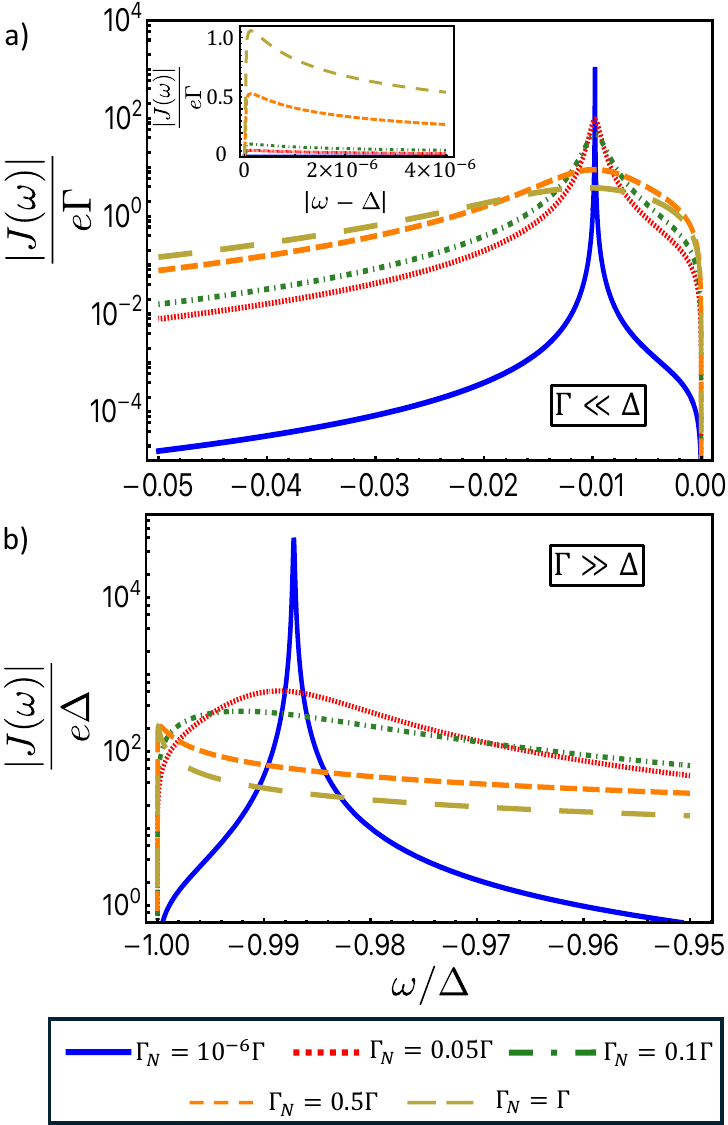}
        \caption{Frequency profile of the Josephson current for different values of the coupling with the normal lead $\Gamma_N$ at fixed phase bias $\phi=0.3 \,\mathrm{rad}$ for the weak- (above (a)) and strong- (below (b)) coupling regimes. All other parameters are as for the respective cases in Fig.~\ref{fig: 3_Jin_Jout_panel} but with $T=0$.}
        \label{fig: 4_Jom_panel}
    \end{figure}

In the numerical simulations, the energies are measured in units of $\Delta$ ($\Gamma$) in the weak (strong) coupling, representing the highest energy scale in that regime. Correspondingly, since the energy scale of the quasi-ABS are of order $\Gamma$ ($\Delta$) in the weak (strong) coupling, the Josephson current is measured in units of $e\Gamma$ ($e\Delta$) (in order to have the maximum current normalized to unity)~\cite{Capecelatro2023}.

In Fig.~\ref{fig: 2_Current_Matsubara_SC} we show the junction CPR at different values of the dot - N lead hybridization in the strong coupling. The overall effect of the normal reservoir is a reduction of the current amplitude and a change of the CPR from a sawtooth shape, typical of ballistic junctions in resonant tunneling, i.e. $\varepsilon_{d}=\mu_{N}=\mu=0$, to a more sinusoidal one, typical of tunnel junctions. 
In other words, at finite $\Gamma_{N}$, the dot level broadening induces decoherence on the system that behaves as if it had a much lower barrier transparency, similarly to what happens when the quantum dot is tuned off resonance with respect to the S leads~\cite{Bagwell1992, Beenakker1992, JonckhereeMartin2009}, see Appendix~\ref{app: epsd neq}.
A similar CPR  characterizes the weak-coupling limit, which is not shown here. 
Moreover, the reduction of the JJ current is visible also in the non-resonant regime, $\varepsilon_{d}\neq0$, Fig.~\ref{Fig_CPR_offres_SC} in Appendix~\ref{app: epsd neq} (strong coupling), that hence show the same phenomenology of the resonant one.

In both weak- and strong-coupling regimes, when $\Gamma_{N}\rightarrow 0$ the current is mostly carried by the Andreev levels. This is evident in Fig.~\ref{fig: 3_Jin_Jout_panel}, where the sub-gap and supra-gap contributions at fixed phase bias $\phi=0.3$ as a function of the $\Gamma_{N}/\Gamma$ ratio are reported.
This phase difference between the S electrodes is chosen in such a way to have the ABS energies close to $\Delta$ in the strong coupling, so as to analyze both the cases in which they lie far and close to the gap. 
 
Interestingly, while in the weak-coupling regime, Fig.~\ref{fig: 3_Jin_Jout_panel}(a), the Josephson current approximately coincides with the sub-gap contribution, this is not true when the interaction between the dot and the S electrodes is larger than $\Delta$, Fig.~\ref{fig: 3_Jin_Jout_panel}(b).     
Nevertheless, in the former case while the sub-gap contribution decreases as $\Gamma_{N}$ grows, the supra-gap current remains constant until $\Gamma_{N}\lesssim\Gamma$ while it is slightly enhanced for $\Gamma_{N}\gg\Gamma$.
 
In the opposite regime, Fig.~\ref{fig: 3_Jin_Jout_panel}(b), not only does the $J^{out}$ decrease with $\Gamma_{N}$, but its decay is faster than that of $J_{in}$, indicating that also the supra-gap current suffers from the decoherence induced by the normal metal.

We can better appreciate the Josephson current suppression by looking at frequency dependence of the current density at definite phase bias $\phi=0.3$ in Fig.~\ref{fig: 4_Jom_panel}. 

Sub-gap and supra-gap current contributions have opposite signs, as already found in similar devices~\cite{Benjamin2007, Meng2009_PRB, Sellier2005, Capecelatro2023}, and $J(\omega)$ vanishes at $\omega=\Delta$, in both weak- and strong-coupling limits (for the weak coupling see the inset of Fig.~\ref{fig: 4_Jom_panel} (a)).
The sub-gap \emph{delta-like} resonances,  
whose position roughly coincides with the frequency of the ABS of the closed JJ ($\Gamma_N=0$), get smoothed by increasing $\Gamma_N$. 
However, in the former regime, the peaks stay at least one order of magnitude higher than the background, even at large $\Gamma_{N}$, whereas in the other one they get completely washed out as $\Gamma_{N}$ approaches $\Gamma$. Here, the sub-gap current density profile becomes more similar to that of a continuum spectrum, possibly suggesting, together with the reduction of the supra-gap current $J^{out}$ (Fig.~\ref{fig: 3_Jin_Jout_panel}(b)), a strong interplay between the supra-gap continuum and sub-gap Andreev levels. As a consequence, the quasi-ABS may not be sufficient to describe the sub-gap current in this regime.   

This is in sharp contrast with the weak-coupling limit, where the supra-gap current appears to be unaffected from the normal reservoir and the $J\left(\omega\right)$ peaks survive even at moderate coupling with the environment. Nevertheless, we observe that the sub-gap current density presents sharp peaks very close to the superconducting gap also in the weak-coupling case, see the inset of Fig.~\ref{fig: 4_Jom_panel} (a), before vanishing at $\omega=\Delta$. These peaks become sizable at large $\Gamma_{N}$ values where their height is of the same order of magnitude of the quasi-ABS peaks.
This effect stems from the hybridization between the phase-dependent supra-gap continuum and the normal lead even when the dot is weakly coupled with the leads. This feature is present also in the dot DOS profile and are analyzed in depth in Section~\ref{sec: dos}.  

These numerical results suggest that the coupling with a N reservoir does not only induce a broadening of the sub-gap levels but also produces a sub-gap continuum contribution, that is especially relevant close to $\Delta$. 
Depending on how far the quasi-ABS energies lie from the gap their overlap with the sub-gap continuum can be more or less pronounced, pointing out a first difference between the weak- and strong-coupling limits. In the former case the sub-gap resonances are pinned at energies $\Gamma \ll \Delta$ and do not hybridize with continuum near $\Delta$. In the latter case, where the quasi-ABS energies can get close to the gap, their mixing with the continuum is higher and ends up in the loss of the resonant peaks as $\Gamma_{N}$ increases.

In order to characterize the interplay between the continuum and the ABS more quantitatively we present in the next sections a thorough discussion of the analytic structure of the dot GF. 
Such approach allows us to distinguish the contributions to the current coming from the quasi-ABS from those related to the continuum.
\section{Decomposition of the retarded Green's function and effective NH description}

In this section, we perform a detailed analysis of the singular part of the dot GF in order to discuss how to extract the Andreev NH Hamiltonian of the JJ and how its eigenvalues, i.e. the quasi-ABS energies, contribute to the supercurrent. This is useful to obtain approximated CPR expressions which depends uniquely on the quasi-ABS properties.
	
	\subsection{Green's function decomposition}
	\label{sec: green_decomp}
	
	\begin{figure}[htb!]
		\includegraphics[width=.46\textwidth]{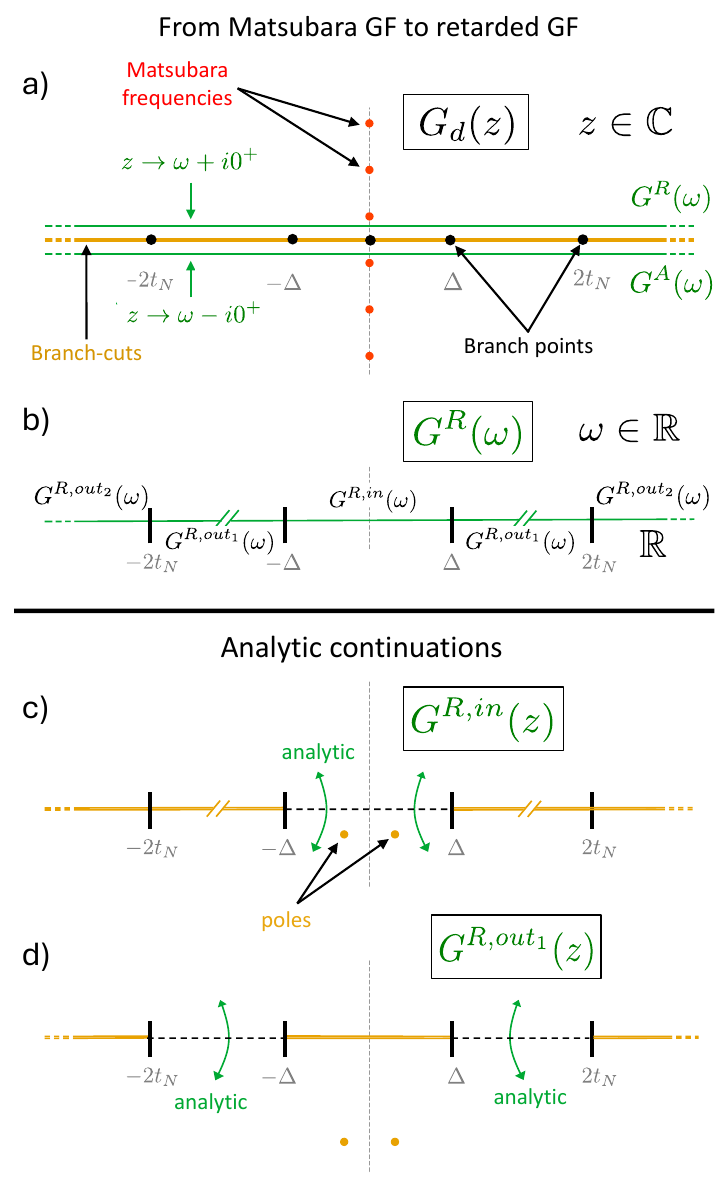}
		\caption{(a) Analytical structure of $\hat G_{d}$ in the complex plane and paths involved in the standard transformation from Matsubara sums to real-frequency integrals of the retarded and advanced Green's functions. (b) Piecewise definition of $\hat G^R$ on the real axis. (c-d) Analytic continuations $\hat G^{R(in/out_1)}$ of the retarded function on the complex plane and their analytic structure.}
		\label{fig-analStr}
	\end{figure}
		
The analytic continuation of the Matsubara Green's function in Eq.~\ref{Gmatsu}, $\hat G_{d}(z)\;(z \in \mathbb{C})$, features square root branch cuts via the S and N leads self-energy contributions. For instance the electron-electron term reads as:

	\begin{eqnarray}
    \label{secular equation}
		\hat G_{d}(z)_{11} &=& \frac{1}{D(z)}\left[z-\varepsilon_{d}+\Gamma\frac{z}{s(\Delta,z)}+\frac{z\Gamma_{N}}{2 t_{N}}\left(\frac{s(2t_N,z)} {s(0,z)}-1\right)\right] \nonumber\\
        D(z) &=&  \epsilon_d^2-\left(z\left(1+\frac{\Gamma}{s(\Delta,z)}\right)+\frac{z\Gamma_{N}}{2t_N}\left(\frac{s(2t_N,z)} {s(0,z)}-1\right)\right)^2 \nonumber\\
        &&+\frac{\Gamma^2\Delta^2}{\Delta^2-z^2}\cos{\left(\frac{\phi}{2}\right)}^2\nonumber\\
	\end{eqnarray}
    
where $s(x,z) = \sqrt{x^2-z^2}$ and $D(z)\equiv\mathrm{det}\left[(\hat G_{d}^{R})^{-1}(z)\right]$. 
For the details about the analytic properties of $\hat G_{d}$ see Appendix~\ref{app: contin}.
Hereafter, since we focus on the dot GF only, we drop the $d$ subscript, for ease of notation.
In order to compute the retarded function $\hat G^R(\omega)$ some care is required, due to the square-root functions, in taking the limit $\hat G(z=\omega+i\eta)$ for $\eta\rightarrow 0^+$. We get:

\begin{equation}
		\label{GRpiecewise}
		\hat G^R(\omega) = 
		\begin{cases}
			\hat G^{R,in}(\omega)  & |\omega| < \Delta \\
			\hat G^{R,out_1}(\omega) & \Delta < |\omega| < 2t_N \\
			\hat G^{Rout_2}(\omega) & |\omega| > 2t_N \\
		\end{cases}
	\end{equation}

See Fig.~\ref{fig-analStr}(b) for a sketch of the functions.
Since the $\hat G^R$ is piecewise defined, we end up having three different analytic continuations $\hat G^{R(s)}(z)$ with $s$ denoting one of the intervals (Fig.~\ref{fig-analStr}(c-d)). 

We focus on the sub-gap region: as soon as the dot - N coupling is turned on, $\hat G^{R,in}$ features a pair of poles with negative imaginary part close to the sub-gap frequency interval. The aim of our analysis, then, is to verify under which conditions in these regimes the GF poles give the main contribution to the observables as in the closed system.
	
We split $\hat G^{R}(z)$ (we drop the apex "in" for the rest of the paper whenever is not strictly needed to lighten the notation) into a singular part, associated to the poles originating from the discrete part of the closed system spectrum, and a branch-cut part~\cite{ColemanBook2015, Zagoskin, Mahan2000}, which is analytic everywhere apart at the branch cuts. 
Therefore, we write
	\begin{equation}
		\hat G^R(z) =   \hat G^{R}_{pol}(z)  +  \hat G^{R}_{bcut}(z).
	\end{equation}
 
As mentioned above, only two simple poles $z_{1,2}$ show up in $\hat G^{R}(z)$, for all parameter regimes, and by the particle/hole symmetry, discussed in Appendix \ref{app: symmetry}, are related by $z_2=-z_1^*$ (see Fig.~\ref{fig:arg_abs_GR} in Appendix~\ref{app: eff_sys_details}). Thus we write polar part of the GF as:

	\begin{equation}
		\label{GR_eff}
		\hat G^{R}_{pol}(z) = \sum_{p=1,2} R_p \hat {P}_p(z - z_p)^{-1}
	\end{equation}

where $R_p\hat {P}_p$ is its residual matrix, equal to $1/(2\pi i)\oint_{\mathcal{C}_p} \hat G^R(z) dz$ with $\mathcal{C}_p$ a small contour around $z_p$. We have split the latter matrix in a rank-1 projector term satisfying $\hat {P}_p^2 = \hat {P}_p$ and $\mathrm{Tr} \hat {P}_p = 1$ and a residual complex number $R_p$. 

The branch-cut GF is analytic at the poles. It shares with $\hat G^{R}$ its branch cuts with their associated discontinuities along the open frequency interval $|\omega|>\Delta$ (see Appendix~\ref{app: bc_funct} for details). 
We stress that $\hat G^{R}_{bcut}$ is always non-vanishing except when the dot is disconnected from all the leads and it does not owe specifically to the coupling with N.

\subsection{Connection between polar Green's function and the effective non-Hermitian system}
	\label{sec: polarAndNH}

In Appendix~\ref{app: eff_sys_details} we show that the polar Green's function in Eq.~\ref{GR_eff} can be rewritten as:

	\begin{equation}
		\label{GR_eff_neq}
		\hat G^{R}_{pol}(z) = \quad
		\hat Z (z - \hat H_{eff})^{-1} 
	\end{equation}

where we defined the non-Hermitian matrix $\hat H_{eff} = \sum_p z_{ p} \,(R_p \hat Z^{-1}\hat {P}_{p})$ and the matrix $\hat Z = \sum_p R_p \hat {P}_p$. They obey particle/hole symmetry constraints $\hat H_{eff} = -\hat \tau_y\,\hat H_{eff}^* \hat \tau_y $ and $\hat Z = \hat \tau_y\,\hat{Z}^* \hat \tau_y $. 

Crucially, these matrices are frequency independent and naturally extend the definitions of renormalized energies and quasiparticle weights in standard Landau theories to a multiorbital case~\cite{Sko22,Bla23}. They are $2\times 2$ matrices acting on the Nambu space, are defined non-perturbatively in any parameter and do not require the broad-band approximation for the normal lead, usually adopted in literature~\cite{Shen2024, Pino2024, Cayao2024, CayaoSato2024, CayaoAguado2024}. 

Systems governed by the polar Green's function and characterized by $\hat H_{eff}$ and $\hat Z$ can be viewed as  \emph{non-Hermitian (NH) systems}. We refer to $\hat H_{eff}$ as the \emph{Andreev} NH Hamiltonian and to $\hat Z$ as its weight matrix. We stress that this effective system does not act on the Hilbert space of the whole junction but only on dot one.  

We say that our model is \emph{approximated} by a effective NH system when the $\hat G^R_{bcut}$ does not produce sizable contributions to the observables at hand.

If $\varepsilon_d = 0$ it is immediate to see that $\hat {P}_1$ and $\hat {P}_2$ are independent of $\phi$ and project orthogonally onto mutually orthogonal eigenspaces of $\hat \tau_x$, since $\hat G^{R}(z)$ commutes with $\hat \tau_x$. This is a consequence of the fact that the system is at half-filling, implying that the poles identify charge-neutral occupation at the dot. The complementarity of the projectors implies that $[\hat Z,\hat H_{eff}]= 0$, making it evident that the eigenvectors of the NH Hamiltonian have also a well-defined residual weight attached to them. In particular we have the simplification $R_p \hat Z^{-1}\hat {P}_{p}= \hat {P}_{p}$. It is also straightforward to see, here, that the system cannot host exceptional points, where the Andreev levels and eigenstates coalesce, since that would imply $\hat {P}_1 = \hat {P}_2$ and, in particular, a lack of mutual orthogonality.

By contrast, at $\varepsilon_d \neq 0$, the projectors $\hat {P}_p$ are not complementary anymore and $[\hat Z,\hat H_{eff}] \neq 0$ (see Appendix~\ref{sec: non_comm}). Still, it is possible to show that there are no exceptional points in the system (see Appendix~\ref{app: EP}). The fact that the eigenvectors of $\hat Z$ and those of $\hat H_{eff}$ do not coincide calls for a deeper understanding of the interplay between the two matrices. Indeed, it challenges the simple picture of the effective system as being driven by a Schr{\"o}dinger equation $i\partial_t \psi = \hat H_{eff}\psi $ for the dot state. We relegate further investigations on this topic to future studies.
Finally, in Appendix~\ref{app: NHredu} we show that that the NH Hamiltonian reduces to a Hermitian one when the N lead is decoupled. In the same limit $Z$ becomes real and proportional to the identity. 

    \section{Poles and Density of States}

    In this section, we analyze both the dot GF poles and the DOS by varying the coupling with the N reservoir. The position and the broadening of the quasi-ABS peaks can be predicted by means of two analytic expressions for the dot GF poles, which are easy to derive in the weak- and strong-coupling regimes. Moreover, by exploiting the results of Section~\ref{sec: green_decomp} we compute the contribution of the quasi-ABS to the DOS that is a measure of the quality of approximating the dot GF with its singular part, $\hat{G}_{pol}^{R}$.
    
    \subsection{Poles analytic expressions}
    \label{sec: poles analytic expression}
     \begin{figure}[hbt!]
		\centering
        \includegraphics[scale=0.6]{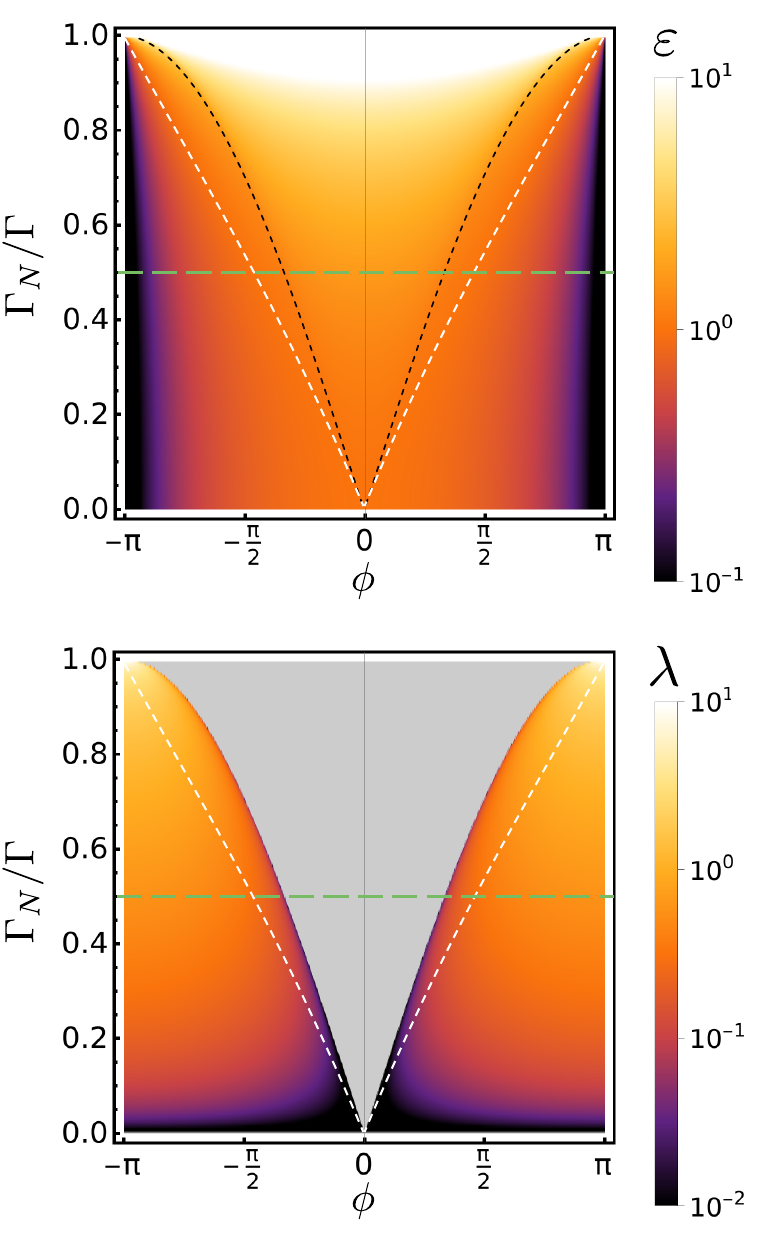}
		\caption{Real part ($\epsilon$, above) and imaginary part ($\lambda$ below) of the quasi-bound Andreev levels in the strong-coupling regime with broadband approximation as per Eq.~\ref{ABS_zBroadB}. The white dashed line shows the phase $\phi_0(\zeta=\Gamma_{N}/\Gamma)$ at which $\epsilon = \Delta$. The black dashed line above shows the phase at which the imaginary part gets quenched. The green lines show the value of $\zeta$ at which Fig.~\ref{fig: Andreev vs Delta} is evaluated. Other parameters (in units of $\Gamma$): $\varepsilon_d = 0, \Delta = 0.01, t_N = 10$.}		\label{fig: z_phi_R_density}
    \end{figure}
        \begin{figure}[hbt!]
		\centering
		\includegraphics[scale=0.4]{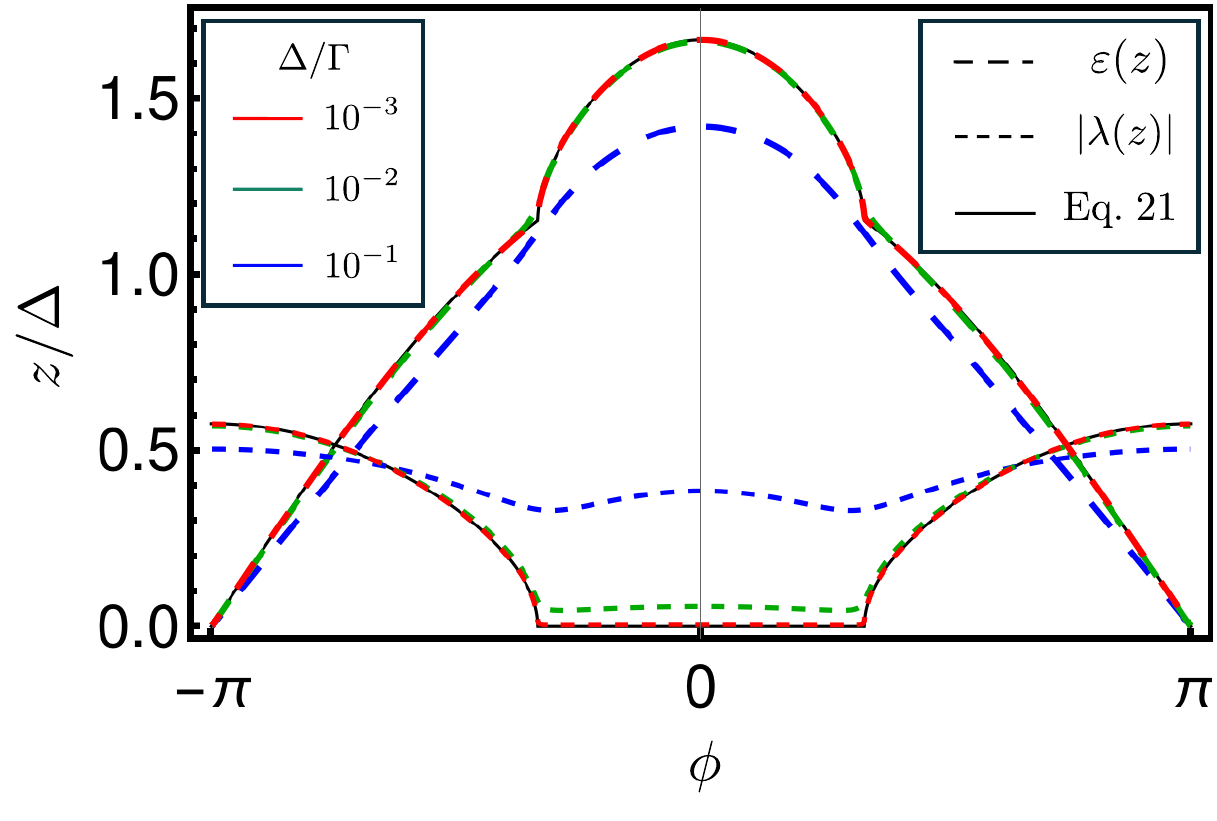}
		\caption{Real part (wide dashed lines) and imaginary part (steep dashed lines) of the quasi-bound Andreev levels in the strong coupling as a function of the phase for different ratios $\Delta/\Gamma$ at fixed ratio $R = \Gamma_N/\Gamma = 0.5$. Other parameters (in units of $\Gamma$): $\varepsilon_d = 0, t_N = 10$.}		\label{fig: Andreev vs Delta}
    \end{figure}

    The poles $z_p$ of the dot retarded GF can be analytically retrieved  at resonance and in the broadband approximation for the normal lead. 
    Setting $D(z) = 0$, see Eq.~\ref{secular equation}, in the weak-coupling regime, we get one pole at
    \begin{equation}
        \label{weak_tunneling_poles}
        z=\sqrt{\varepsilon_{d}^2+\Gamma^{2} \cos{\left(\phi/2\right)}^{2}} - i \Gamma_{N}\,
    \end{equation}
    while the other one can be retrieved by particle/hole symmetry. 
    Its real part coincides with the well-known formula for the Andreev levels of the closed QD JJ with Breit-Wigner transparency~\cite{Beenakker1991_B,Beenakker1992,Beenakker2024} and is unaffected by the coupling with the normal lead. The imaginary part simply coincides with the dot--N lead self-energy term. As a consequence, the quasi-ABS have the same phase-independent broadening that the dot level has when coupled to the normal metal only. Therefore, the effects of the coupling of the QD to the S and N leads are independent: the former is responsible for the change in the real energy of the quasi-ABS while the latter of their line-widths.
    
    In the strong-coupling limit, for $\Gamma_{N}<\Gamma$ and $\Delta\ll\Gamma$, the pole is given by a cumbersome expression. At $\varepsilon_{d}=0$ it simplifies to
    \begin{equation} \label{ABS_zBroadB}
        \begin{split}
            z=&\frac{\Delta}{1-\zeta^2}\cos\left(\phi/2\right)-\\
            &-i\frac{\Delta\, \zeta \, \mathrm{sgn}(\cos\left(\phi/2\right))}{\left(1-\zeta^2\right)}\sqrt{\sin{\left(\phi/2\right)}^2-\zeta^2}\,,
        \end{split}
    \end{equation}
    where we defined the ratio $\zeta=\Gamma_{N}/\Gamma$.

   The analytic expression in Eq.~\ref{ABS_zBroadB} is plotted in Fig.~\ref{fig: z_phi_R_density} against both $\zeta$ and $\phi$ and in Fig.~\ref{fig: Andreev vs Delta} we show it to be a perfect approximation of the numerical GF poles only when $\Delta \lesssim 10^{-3} \Gamma$.

    In the closed junction limit the formula gives the Andreev levels energy for a single-channel short-junction in resonant tunneling~\cite{Beenakker1991_A, Beenakker1991_B,Beenakker1992, Beenakker2024,Li2024,Golubov2004, Furusaki_Takayanagi_Tsukada_1992}. In the open junction case, the imaginary part acquires a non-trivial phase dependence, implying that the effective broadening of the ABS, $\Gamma_{eff}$, is affected by the junction phase bias. Moreover, $\Gamma_{eff}$ is proportional to the factor $\Delta \,\zeta/\left(1-\zeta^2\right)$ which can be very small at small $\zeta$ but gets rapidly enhanced as its value increases. Importantly, at fixed value of $\zeta$, for phases in an interval $[-\phi_0(\zeta),\phi_0(\zeta)]$, the real part of the poles lies outside the sub-gap region and, in an even smaller interval, it has vanishing imaginary part, as indicated respectively by the white and black dashed lines in Fig.~\ref{fig: Andreev vs Delta}). Lifting the constraint $\Delta\ll\Gamma$, the imaginary part never vanishes but a significant quench remains, see Fig.~\ref{fig: Andreev vs Delta}. We checked that, changing the superconducting leads self-energy to a more accurate one, describing them as 1D chains~\cite{Furusaki1994}, does not alter qualitatively the presence of a non-trivial curve $\phi_0(\zeta)$.
    The fact that the real part of the GF poles moves outside the superconducting gap does not correspond to any specific feature in the CPR, see the orange short-dashed curve in Fig.~\ref{fig: 2_Current_Matsubara_SC}. This questions the link between GF (or equivalently the scattering matrix) poles and the observable current, possibly suggesting the incompleteness of a quasi-ABS picture in this transport regime.

    \subsection{The Density of States}
    \label{sec: dos}
    The dot DOS is defined as
	\begin{equation}
		\label{Dos_QD}		
  \rho\left(\omega\right)=-\frac{1}{\pi}\mathrm{Im\,Tr}\,\hat G^{R}(\omega) 
	\end{equation}
and it is normalized so as to have $\int_{-\infty}^{\infty}\mathrm{d}\omega \,\rho(\omega)=2$.

Resonant structures are indeed expected to be observable in the frequency profile of the quantum dot DOS in correspondence of the Andreev sub-gap states~\cite{Furusaki_Tsukada_1991}.

    \begin{figure}[hbtp]
        
        \includegraphics[width=0.48\textwidth,height=0.68\textwidth]{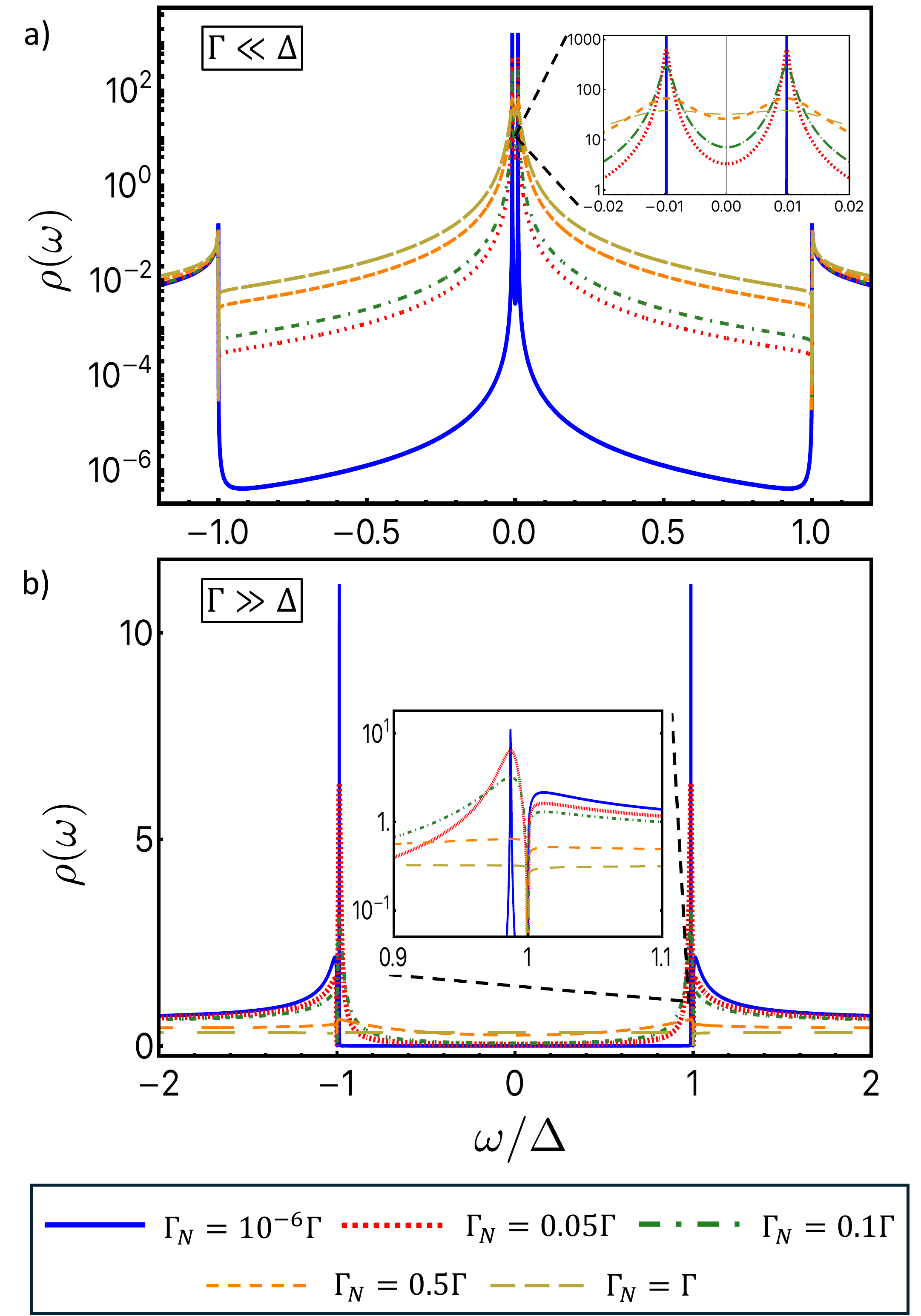}
        \caption{Frequency profile of the dot density of states at the gap for different values of the coupling with the normal lead $\Gamma_N$ at fixed phase bias $\phi=0.3 \,\mathrm{rad}$ for the weak (above (a)) and strong (below (b)) coupling regimes.  All other parameters are as for the respective cases in Fig.~\ref{fig: 3_Jin_Jout_panel}, i.e. $\varepsilon_{d}=0$, $t_{N}=10$ and $\Gamma=0.01$ (in units $\Delta$) in the weak coupling and 
        $\varepsilon_{d}=0$, $t_{N}=10$ and $\Delta=0.01$ (in units $\Gamma$) in the strong coupling.}
        \label{Fig_Rho_Weak_Strong_Panel}
    \end{figure}
In Fig. \ref{Fig_Rho_Weak_Strong_Panel} we  show the DOS computed numerically in both regimes. 

At weak coupling, Fig.~\ref{Fig_Rho_Weak_Strong_Panel} (a), the ABS resonances occur at energy scales of $\Gamma$ as predicted by Eq.~\ref{weak_tunneling_poles}. By contrast, at strong coupling, Fig.~\ref{Fig_Rho_Weak_Strong_Panel} (b), the peaks shift very close to the superconducting gap $\Delta$ at small phase bias. 
By increasing $\Gamma_{N}$ these resonances get lower and wider in all regimes. However, in the former case the height of the peaks stays much larger than the background continuum spectrum as it happens for the current density. This happens even at $\Gamma_{N}\sim\Gamma$, thus testifying a negligible overlap between quasi-ABS and the continuum states. In the latter regime, the sub-gap resonances, whose spectral weight is much lower than the supra-gap states, are completely washed out as $\Gamma_{N}$ approaches $\Gamma$.

The proximity of the quasi-ABS to $\Delta$ hints at a high spectral mixing with the supra-gap states when their broadening is sizable.
Further, we note that the curves computed at $\Gamma_{N}$ equal to $0.5\Gamma$(dashed orange) and $\Gamma$(wide-dashed yellow) in Fig.~\ref{Fig_Rho_Weak_Strong_Panel} refer to situations in which ABS peaks are no more visible since their energy (the real part of the GF poles) has moved outside superconducting gap. This suggests that a sizable sub-gap contribution to the DOS can be ascribed to the continuum.

In order to analyze the interplay between quasi-bound states and the continuum in this regime, we exploit the decomposition of the retarded function, by splitting the sub-gap DOS profile into two terms, the polar and the branch-cut one
		\begin{equation}
            \label{dos_splitting}
			\rho^{in}(\omega) = -\frac{1}{\pi}\mathrm{Im\,Tr}\,\hat G^{R,in}(\omega) =\rho_{pol}^{in}(\omega) + \rho_{bcut}^{in}(\omega) \,.
		\end{equation}

		It can be proved easily, using the representation Eq.~\ref{GR_eff} of the polar retarded function, that 
	\begin{eqnarray}
			\rho_{pol}^{in}(\omega) &=& R^r \left(L_{\epsilon,\lambda}(\omega) + L_{-\epsilon,\lambda}(\omega) \right) - \nonumber\\
			&& -
			R^i  \left(\frac{\omega-\epsilon}{\lambda}L_{\epsilon,\lambda}(\omega) - \frac{\omega+\epsilon}{\lambda}L_{-\epsilon,\lambda}(\omega) \right) 
		\end{eqnarray}
		where $ L_{\epsilon,\lambda}(\omega) = \frac{1}{\pi} \frac{\lambda}{(\omega - \epsilon)^2 + \lambda^2}$ is a Lorentzian function and we associated the residue $R=R^r + i R^i \;(R^{r/i}\in \mathbb{R})$ to the pole at $\epsilon - i \lambda$.
		
	   The polar part of the DOS, associated to the quasi-ABS, describes a Fano resonance typical of discrete levels coupled to a continuum~\cite{Mir10,Erd17}. This is made of two qualitatively different contributions: a Lorentzian centered at the real part $R^r$ of the poles plus a contribution which is odd with respect to the Lorentzians centers and proportional to the imaginary part $R^i$ of the residues at the poles. $R^i$ stems directly from the coupling of the junction with the normal lead, see Appendix~\ref{app: NHredu}.
  
    Some details about the form of the weight matrix $\hat Z$ and the residue in the two coupling regimes can be found in Appendix~\ref{app: ZandRes}.

\begin{figure}[hbtp!]
		\centering
		\includegraphics[scale=0.59]{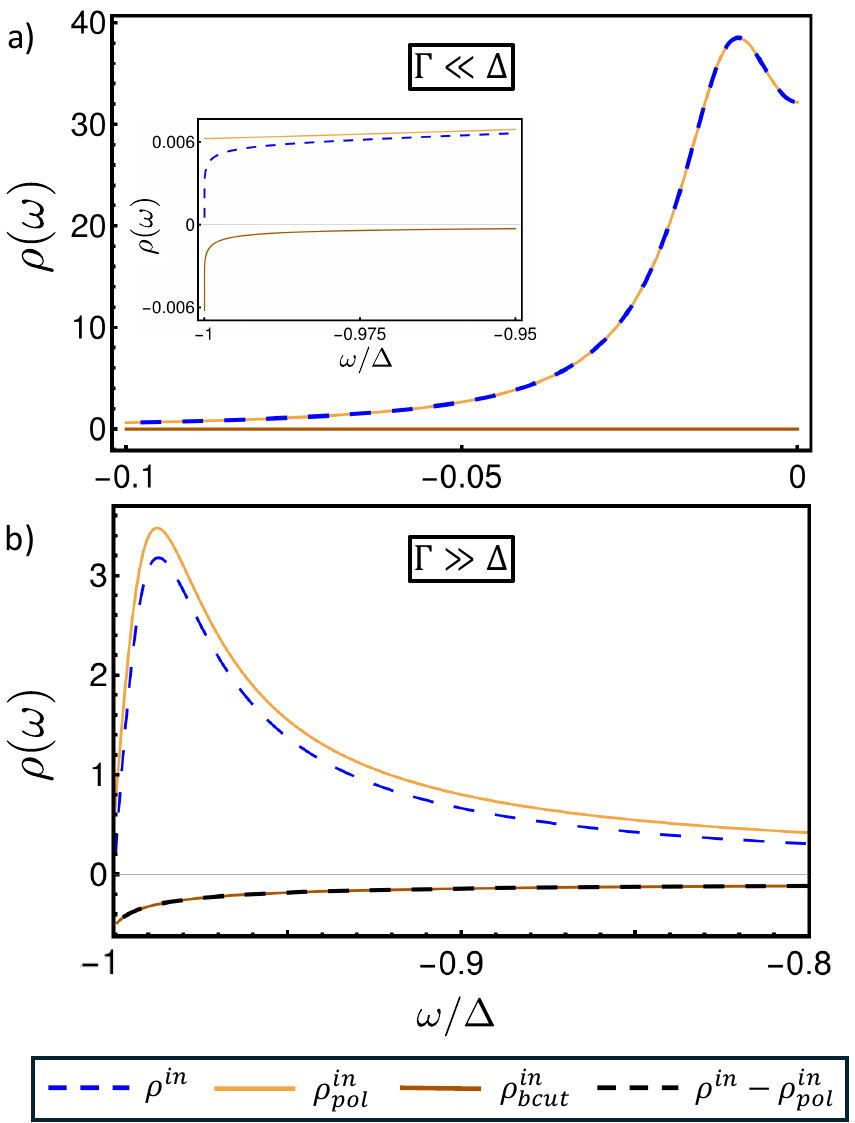}
		\caption{Density of states and its components in the weak- (a) and strong- (b) coupling regimes. All curves are obtained using the definition in Eq.~\ref{Dos_QD} but different retarded functions. $\rho$ is computed using $\hat G^R$, $\rho_{pol}^{in}$ using $G^{R,in}_{pol}$ and $\rho_{bcut}^{in}$ using $G^{R,in}_{bcut}$ in Eq.~\ref{Gbcut_formula}. In the lower panel, we plot the difference of the first two 
        densities as a numerical check of the expression Eq.~\ref{Gbcut_formula}. The respective parameters are as in Fig.~\ref{Fig_Rho_Weak_Strong_Panel}, with $\Gamma_N = \Gamma$ (a) and $\Gamma_N = 0.1 \Gamma$ (b). }		\label{fig: rho_bcut_check}
    \end{figure}

  Clearly, $\rho_{bcut}^{in}\equiv 0$ in the closed system since there cannot be any continuum spectral weight. 
  In the weak-coupling case, $\rho_{bcut}^{in}$ can be disregarded, see Fig.~\ref{fig: rho_bcut_check} (a). At $\omega\simeq\Delta$, see inset in Fig.~\ref{fig: rho_bcut_check} (a), there is a increased contribution but becomes sizable only when $\Gamma_{N}\gg\Gamma$ and its overlap with $\rho_{pol}^{in}$ is small since the quasi-ABS stay far from the gap. 

  On the contrary, in the strong-coupling regime, $\rho_{bcut}$ becomes relevant at moderate values of $\Gamma_{N}$, especially at small phase biases when the quasi-ABS are close to the gap edge. This contribution overlaps with the near-gap quasi-ABS affecting the typical Fano-shape of the resonance, see  Fig.~\ref{fig: rho_bcut_check} (b).
  
  An intuitive reason for this increased $\rho_{bcut}$ contribution can be traced back to a fundamental property of the system. Indeed, one may observe that the branch-cut contribution enforces the vanishing of the DOS at $\omega = \Delta$.
  Such peculiar behavior is inherited by the dot GF from the one of the S leads which feature a Van Hove singularity at the gap~\cite{Mahan2000, ColemanBook2015, Zagoskin} and it is not imprinted in the Fano resonance itself. In other words, when the quasi-ABS resonances are found in the proximity of the gap, their spectral distribution $\rho_{pol}^{in}$ would not automatically vanish at $\omega=\Delta$. Then, a branch-cut contribution with a negative sign must emerge to balance the finite quasi-ABS spectral weight and is larger the higher is the value of $\rho_{pol}^{in}(\Delta)$.

    \begin{figure}[hbtp]
		\centering
		\includegraphics[scale=0.25]{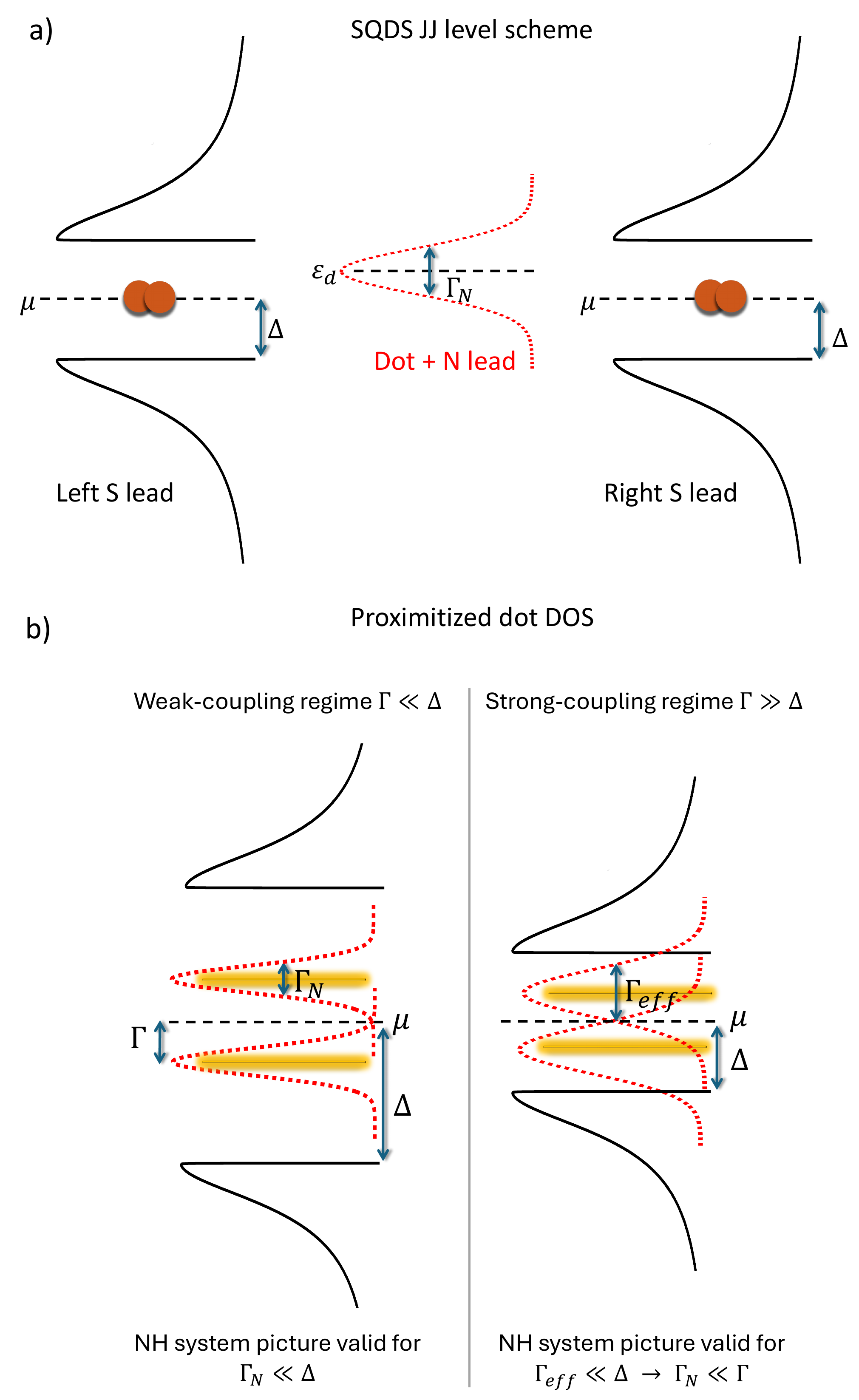}
		\caption{In the upper panel (a) we show a sketch of the SQDNS JJ levels. The superconductor are represented as a Cooper pair condensate that is separated by the continuum quasiparticles excitations by $\Delta$. The quantum dot level, whose energy is $\varepsilon_{d}$ and its line-width is $\Gamma_{N}$, is broadened by the influence of the N - lead (red dashed line). In the lower panel (b) we report a sketch of the proximitized quantum dot DOS for an open junction in the weak- and strong-coupling limits. In the first case, the ABS (in yellow) occur at energy scales of $\Gamma\ll\Delta$ while their broadening is given by the normal lead - dot hybridization $\Gamma_{N}$ (red dashed line). In the second situation, the ABS energy is at energy scales of $\Delta\ll\Gamma$ and their imaginary part can be interpreted as an effective broadening $\Gamma_{eff}$ (red dashed line) that induces a overlap between the sub-gap and the supra-gap states.}
        \label{Fig_Sketch_SQDNS_Panel}
    \end{figure}
     To summarize, we sketch the interplay between the poles and the continuum in the DOS in the two regimes in Fig. \ref{Fig_Sketch_SQDNS_Panel}. 
     
     In the weak-coupling regime, where the quasi-ABS energy scale is the dot--S hybridization $\Gamma$, these states are always far from the superconducting gap and they inherit the same line-width, $\Gamma_{N}$, of the dot coupled to N, see Fig. \ref{Fig_Sketch_SQDNS_Panel} (a) and (b, left). Thus, the overlap between them and the continuum spectrum is negligible as long as $\Gamma_{N}\ll\Delta$. Here, the sub-gap resonances in the DOS, $\rho\left(\omega\right)$ in Fig. \ref{Fig_Rho_Weak_Strong_Panel} (a), and current density, $J\left(\omega\right)$ in \ref{fig: 4_Jom_panel} (a), survive. 
     
     Differently, in the strong-coupling regime, the quasi-ABS energy scale is of order $\Delta$ and their effective broadening is   
    \begin{equation}
    \label{Gamma_eff}
        \Gamma_{eff} \sim \Delta \zeta/\left(1-\zeta^2\right) 
    \end{equation}
     when ignoring its phase dependence, see Fig. \ref{Fig_Sketch_SQDNS_Panel} (b, right).
     This can get high when $\Gamma_{N}\sim\Gamma \;\;(\zeta\sim 1)$ leading to a severe broadening and a reduction of the quasi-ABS resonances in both the DOS, Fig.~\ref{Fig_Rho_Weak_Strong_Panel} (b), and the current density, Fig.~\ref{fig: 4_Jom_panel} (b).
    We identify three different scenarios: one in which the quasi-ABS are far from the gap (at large phase-bias) where the $\rho^{in}$ is mainly described by $\rho_{pol}^{in}$ as in the weak coupling; one in which the real part of the dot GF poles lie outside the gap, and the major sub-gap spectral contribution is provided by the continuum (see orange and yellow curves in the inset of Fig.~\ref{Fig_Rho_Weak_Strong_Panel}(b)); lastly, one where the quasi-ABS are close to the gap, see Fig.~\ref{Fig_Sketch_SQDNS_Panel} (b, right), with a competition between the poles and the branch-cut DOS contributions.
    \subsection{Requisites for the success of the NH system approximation}

     Following from above considerations, we argue that the Andreev NH approximation is accurate at a generic phase bias as long as the broadening of the quasi-ABS is much smaller than their distance from gap edges. 

    Assuming that the quasi-ABS lie inside the gap, such requirement reads as
    \begin{equation}
    \label{working_GNs}
          \begin{cases}
                \Gamma_{N}\ll\Delta -|\varepsilon_A| \sim \Delta& \;\;  (\Gamma\ll\Delta)\\
                \Gamma_{eff}\ll\Delta -|\varepsilon_A|\overset{(Eq.~\ref{Gamma_eff})}{\rightarrow} \Gamma_{N}\ll \Gamma &\;\;  (\Gamma\gg\Delta) \,.
            \end{cases}
    \end{equation} 
    Instead, when the quasi-ABS resonances overlap with the supra-gap continuuum the branch-cut contribution as to be taken into account to predict the exact density profile, see Fig.~\ref{Fig_Sketch_SQDNS_Panel} (b, right).
    
    We expect that this claim generalizes to any other system in which the 
    barrier single-particle excitations are spectrally well-separated from the continuum spectrum.

    \section{Josephson current from the polar GF and from the effective NH Hamiltonian}
	\label{sec: JPolar and JHeff}
    In the previous section, from the analysis of the spectral contribution to the dot DOS, we identify the relative position of the quasi-ABS with respect to their line-width as a potential fingerprint to validate the Andreev NH description of the JJ. In this section, we test the quality of this argument by computing the current from the Andreev NH Hamiltonian and comparing the results with the prediction of the Green's function formalism, in Section~\ref{sec_3_Transport_Properties}.
        
    \subsection{Current formulas from the polar Green's function}
    \label{sec: Jpol and JHeff formulas}

	To derive the current we consider an expression for free energy of the dressed dot, see Eq.~\ref{Free_energy} in Appendix~\ref{Current in the SQDNS JJ}, as a real-frequency integral~\cite{Benjamin2007, Rozhkov1999}
	\begin{eqnarray}
		\mathcal{F}^{in} = \frac{1}{2\pi i} \int_{-\infty}^{\infty}\mathrm{d}\omega\; n_F(\omega)   \mathrm{Im} \ln \det(\left(G^{R,in}\right)^{-1})
	\end{eqnarray}

We focus on the sub-gap integral and look at the contribution from the polar part of the Green's function. Using Eq.~\ref{GR_eff_neq} we get  
	\begin{eqnarray}
		\mathcal{F}^{in}_{pol}  &=& -\frac{1}{2\pi i} \int_{-\Delta}^{\Delta} \mathrm{d}\omega\; n_F(\omega)   \mathrm{Im} \ln \left(\det \hat Z (\omega - \hat H_{eff})^{-1}\right) \nonumber\\ 
		&=& -\frac{1}{2\pi i} \int_{-\Delta}^{\Delta} \mathrm{d}\omega\;  n_F(\omega)   \Big[\mathrm{Im} \ln \left(\det \hat Z \right) + \nonumber\\
		&& + \mathrm{Im} \ln \left(\det (\omega - \hat H_{eff})^{-1}\right)\Big]   \nonumber\\ 
		&=&  \frac{1}{2\pi i} \int_{-\Delta}^{\Delta}  \mathrm{d}\omega\;  n_F(\omega) \mathrm{Im} \ln \left(\omega - z_1\right)\left(\omega - z_2)\right)   \nonumber\\ 
        \label{Polar_Free_energy}
	\end{eqnarray}
	Here, the weight matrix in the second row disappears because its determinant is real, due to the particle/hole symmetry (see Appendix~\ref{app: ZandRes}). As a consequence, in regimes where the effective NH system approximation is found valid, from the point of view of the current, there is no difference whether it is computed from our system or one with $\hat Z = \hat 1$, i.e. a system driven by a pure NH Schr{\"o}dinger equation. 
    Clearly, the same claim holds for any other observable which derives of the free-energy, so long as particle/hole symmetry is preserved by the external (static) fields.

    It is easy to derive the current using the known relation $J =-2e \partial_\phi \mathcal{F}$\cite{Beenakker1992} and valid also in the open junction\cite{Beenakker2024,Pino2024}.
    Recalling that $z_{1,2}=\pm\varepsilon-i\lambda$, we get 
	\begin{eqnarray}
		\label{Jpol_general}
		J^{in}_{pol}(\phi)/2e &=&  -\frac{1}{\pi} \Big\{ (\partial_\phi \epsilon)  \int_{-\Delta}^{\Delta} \mathrm{d}\omega\; \frac{\tanh(\beta\omega/2)\,\lambda}{\left[(\omega + \epsilon)^2 + \lambda^2\right]} + \nonumber\\
		&& - (\partial_\phi \lambda)  \int_{-\Delta}^{\Delta} \mathrm{d}\omega\; \frac{\tanh{(\beta\omega/2)}(\omega + \epsilon)}{\left[(\omega + \epsilon)^2 + \lambda^2\right]} \Big\}. \nonumber\\
	\end{eqnarray}
	The first contribution is simply a Lorentzian-broadened quasi-ABS contribution, while the second term is entirely due to the non-Hermiticity and resembles the asymmetric contribution of the Fano resonance. This term gets larger the more the quasi-ABS lifetimes are sensitive to the phase and the more its real energy is close to the gap edge. This formula generalizes the known result at $\Gamma_N=0$,  $J^{in}_{pol} = -2e(\partial_\phi \epsilon) \tanh(\beta\epsilon/2)$\cite{Beenakker1992}.
	At zero temperature the formula reduces to
	\begin{eqnarray}
		\label{Jpol_T0}
		J^{in}_{pol}(\phi)/2e &\overset{T\rightarrow0}{=} &  -\frac{\partial_\phi \epsilon}{\pi} \Big\{ 2\arctan\left(\epsilon/\lambda\right) - \nonumber\\ 
		&& -\arctan\left[2\left(\epsilon \lambda\right)/(\Delta^2+\lambda^2-\epsilon^2)\right] \Big\} +\nonumber\\
		&& + \; \frac{\partial_\phi \lambda}{2\pi} \ln\left( \frac{|z_1|^4}{(|z_1|^2-\Delta^2)^2 +4\lambda^2\Delta^2}\right)\,.
		\nonumber\\
	\end{eqnarray}
 
    The above equation applies to all transport regimes and reduces to the result in Ref.~\cite{Beenakker2024} when taking the limit $\Delta \rightarrow \infty$ and setting $\lambda$ constant.
	
	In Ref.~\cite{Shen2024} a different formula for the current has been obtained which applies to our geometry. The formula is a function of the effective NH Hamiltonian of the whole junction $H^{JJ}_{eff}$, where only the normal lead degrees of freedom are integrated out. As such, it includes both sub- and supra-gap currents. At $T=0$ reads as $J(\phi)/2e = -\frac{1}{\pi}\partial_\phi\,\mathrm{Im} \,\mathrm{Tr} \left(H^{JJ}_{eff} \ln H^{JJ}_{eff}\right)$.  In the following section we compare the the current from Eq.~\ref{Jpol_general} with the one predicted from this formula inserting our $2\times 2$ matrix $\hat H_{eff}$, instead of the much bigger $H^{JJ}_{eff}$. This simplification has been proved to be faultless in the weak-coupling limit in Ref.\cite{Beenakker2024}. Our version of the formula reads as
	\begin{eqnarray}
		\label{Jshen_lado}
		J^{in}_{H_{eff}}(\phi)/2e &\overset{T\rightarrow 0}{=}& -\partial_\phi \epsilon \,\left( \mathrm{sgn}(\epsilon) -\frac{2}{\pi}\arctan\left(\frac{\lambda}{\epsilon}\right)\right) + \nonumber\\ 
		&& + \frac{2}{\pi}\,\partial_\phi \lambda\,\left(\ln |z| +1\right)\,
	\end{eqnarray}

    and coincides with Eq.~\ref{Jpol_T0} upon taking the limit $\Delta \rightarrow \infty$ and setting $\lambda$ to a constant.
    
	\subsection{Polar current at different regimes}
    \label{sec: Jpol and JHeff numerics}
    \begin{figure*}[htbp]
		\centering
		\includegraphics[width=1\linewidth]{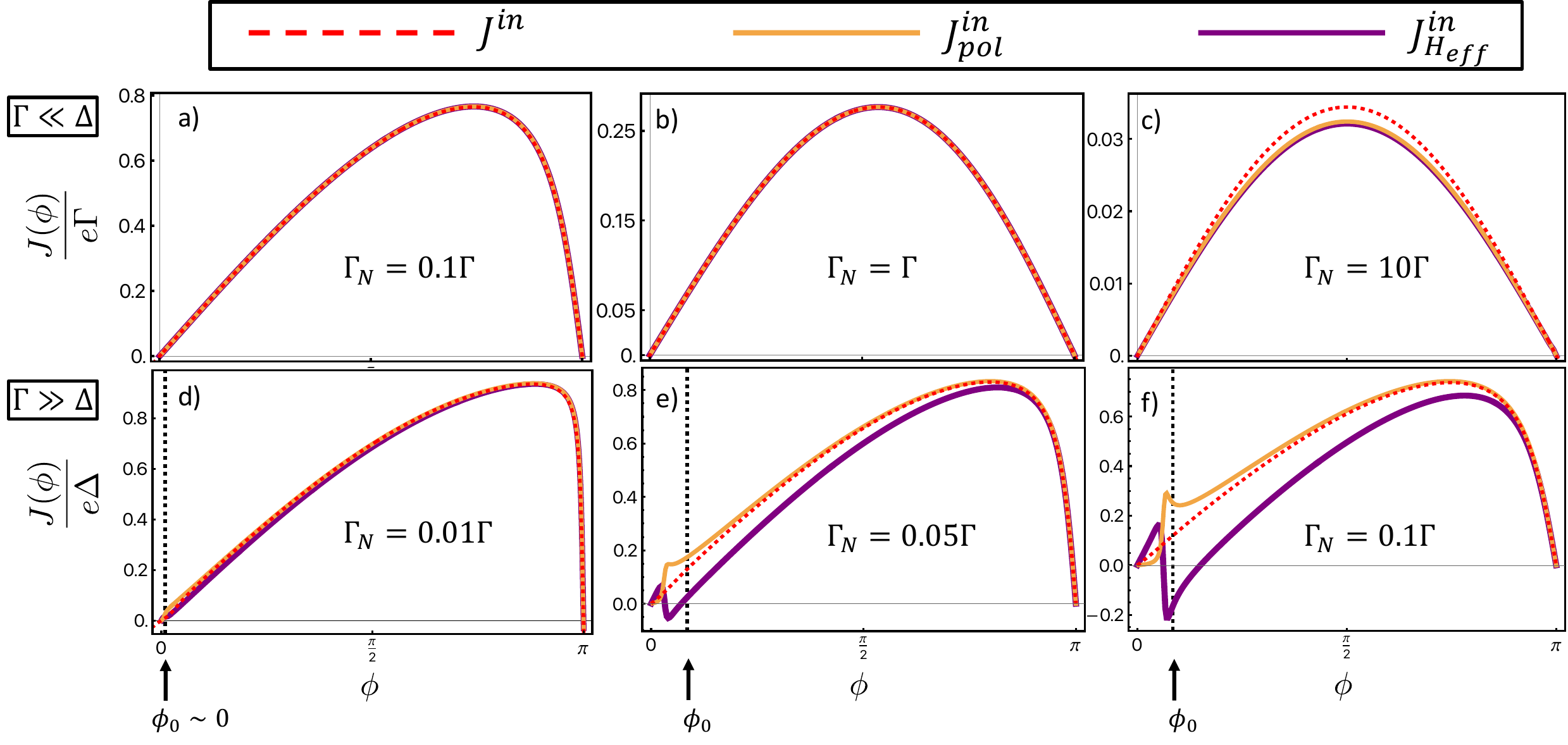}
		\caption{Performance of current formulas in Eqs.~\ref{Jpol_T0}-\ref{Jshen_lado} in the weak- (first row) and strong- (second row) coupling regimes. $J^{in}$ is the exact current, see Eq.~\ref{Andreev levels current}, taken as reference. $J^{in}_{pol}$ is taken from Eq.~\ref{Jpol_general} while $J^{in}_{H_{eff}}$ from Eq.~\ref{Jshen_lado}. All curves are at $T=\varepsilon_d = 0$. The other parameters are as for the respective cases in Fig.~\ref{fig: 4_Jom_panel}.}
		\label{fig: J_compar}
	\end{figure*}
	Here, we investigate numerically if the total current can be approximated by Eqs.~\ref{Jpol_general} and \ref{Jshen_lado} in the two regimes of weak and strong coupling, separately. 
	
	As it can be seen from Fig.~\ref{fig: J_compar}(first row), in the weak-coupling regime the sub-gap current from the NH system approximation, Eq.~\ref{Jpol_general}, is in excellent agreement with exact one from Eq.~\ref{Andreev levels current}. We see a failure of the polar current formula only at very high values of the coupling with the lead.
 Yet, the mismatch at $\Gamma_N = 10 \,\Gamma$ (Fig.~\ref{fig: J_compar} (c)) is surprisingly high, with respect to what we expected from the considerations in Section~\ref{sec: dos}. 
 This failure is related to the sharp peak in the current density close to $\omega=\Delta$, shown in the inset of Fig.~\ref{fig: 4_Jom_panel}, which cannot be described within the NH system approximation. This peak is related to the presence of a sub-gap branch-cut current contribution at $\omega\sim\Delta$.
 This implies that a better match could be found by reducing the frequency sub-gap window to exclude that peak. We explain this in details in Appendix~\ref{Fail_Pol_weak_coupling}
 
    Also the current from Eq.~\ref{Jshen_lado} exactly matches with the results from Eq.~\ref{Andreev levels current}. 

    One can check that the NH Hamiltonian computed from Eq.~\ref{GR_eff_neq} coincides with the one showed in Ref.\cite{Beenakker2024}, evaluated at equal right and left hybridization factors.

	In the strong-coupling regimes the picture is more complex, see Fig.~\ref{fig: J_compar}(second row). For $\Gamma_{N} \ll \Delta$ all current formulas are in excellent  agreement except for small phases.  Near the phase $\phi_0$, where the real energy of the poles exits the gap, we observe a strong deviation of the approximated sub-gap currents with respect to the exact one.
    The failure is harsher the larger is the coupling. We mention that $J_{H_{eff}}$ performs systematically worse than $J_{pol}$. Indeed, formula \ref{Jshen_lado} was conceived to apply to the Hamiltonian of the whole JJ, including the S leads degrees of freedom, and assuming no retardation effects from the N-lead self-energy. Instead in our case, the S leads self-energies bring a retardation which, in the strong-coupling regime, has a sizable effect.
    
    To summarize, the current observable seems to be reliably approximated by the one of the effective NH system only in the weak-coupling regime provided that the contribution of the peak at $\Delta$ is negligible. In the strong-coupling regime only currents at phases far from $\phi_0$ can be approximated. These results are in agreement and complement what is shown for the DOS in Section~\ref{sec: dos}. In Appendix~\ref{app: epsd neq} we show that detuning the dot off resonance, setting $\varepsilon_d$ to a finite value, does not alter the picture illustrated above. Surprisingly enough, in the strong-coupling regime the NH system approximated current makes better predictions than those at resonance. Indeed, even though the quasi-ABS resonances are pushed at higher values (in modulus), their effective broadening appear strongly reduced, with a net smaller interplay among them and the continuum states.
	
	\section{Summary and Conclusions}

    \begin{table*}[ht!]
    \centering
        \renewcommand{\arraystretch}{1.6} 
        \setlength{\tabcolsep}{4.7 pt}     
        \begin{tabular}{lccccc}
          & \textbf{ABS energy $\varepsilon_{A}$} & \textbf{quasi-ABS} & \textbf{quasi-ABS at $\varepsilon \gtrsim \Delta$} & \textbf{Sub-gap continuum} & \textbf{Success of Andreev NH approach}\\   &&\textbf{broadening}& &&\textbf{(i.e. $\rho^{in}\simeq\rho^{in}_{pol}$ and $J^{in}\simeq J^{in}_{pol}$)} \\ 
         \hline
         \textbf{$\Gamma \ll \Delta$} & $\varepsilon_{A} \sim \Gamma$  & $\Gamma_{N}$  & No  & for $\Gamma < \Gamma_{N} < \Delta$  & $\Gamma_{N} \ll \Delta - |\varepsilon_{A}| \sim \Delta$  \\ 
         \textbf{$\Gamma \gg \Delta$} & $\varepsilon_{A} \sim \Delta$ & $\Gamma_{eff}(\phi)$  & for $\phi \in \left[-\phi_{0}, \phi_{0}\right]$  & for $\Gamma_{eff} \sim \Delta - |\varepsilon_{A}|$  & for $\phi \notin \left[-\phi_{0}, \phi_{0}\right]$ and $\Gamma_{eff} \ll \Delta - |\varepsilon_{A}|$ \\ 
    \end{tabular}
    \caption{Summary of the results obtained from the analysis of the GF poles and the DOS in Secs.~\ref{sec: poles analytic expression}-\ref{sec: dos} and from the comparison between the NH current Eq.~\ref{Jpol_T0} and the results of the exact GF formalism in Section~\ref{sec_2_Model}. In the weak-coupling regime (first row) where the ABS energy are of the order of $\Gamma\ll\Delta$ the broadening is simply provided by the dot-N hybridization, there is no overlap between the quasi-ABS and the continuum near the gap which becomes sizable only for $\Gamma_{N}>\Gamma$. The NH approach perfectly works as long as $\Gamma_{N}\ll\Delta$. In the strong-coupling regime (second row), the ABS lie close to $\Delta$. The coupling with the N lead induce a phase dependent broadening and an energy shift of the ABS which can also exit the gap in a specific phase window $\left[-\phi_{0}(\Gamma_{N}),\phi_{0}(\Gamma_{N})\right]$ . Note that $\phi_{0}(\Gamma_{N})$ grows as $\Gamma_{N}$ increases. More generally, whenever the effective line-width of the quasi-ABS, $\Gamma_{eff}$, is comparable to its distance from $\Delta$, a sizable sub-gap continuum contribution arises. In both these situations the NH approach fails in predicting the exact transport observable.}
    \label{tab:Summary}
    \end{table*}
We considered a JJ with a single-level quantum dot barrier coupled to a normal metal lead. The coupling with the N lead produces a broadening of the ABS encoded in the imaginary part of the dot GF poles.
We proposed a NH Hamiltonian for the quasi-ABS whose accuracy has been tested by benchmarking the transport observables against exact results obtained in the GF formalism. 
We explored the weak- and strong-coupling regimes 
in which the hybridization of the dot with the S leads is, respectively, much lower and larger than the superconducting gap $\Delta$.

The applicability of the NH approach is assessed via analysis of the DOS profile of the dot and is tested by computing the CPR with a novel NH supercurrent formula that we derived. 

The approach is found to work only when the quasi-ABS do not overlap with the continuum states, localized near the gap edges in the DOS profile.
In the weak-coupling regime the quasi-ABS resonances are centered far away from $\Delta$ and the overlap is typically tiny, making the approach excellent. 
In contrast, the approach is much less robust in the strong coupling, where the states always get close to the gap edges at small phase biases. 

In addition to this analysis, we explored whether the small-sized Andreev NH Hamiltonian could be used in the formula of the Ref.~\cite{Shen2024} in place of the full-junction NH Hamiltonian. Such formula performs excellently in the weak-coupling limit but is worse than ours in the strong-coupling one.
Our findings are summarized in Tab.~\ref{tab:Summary} where we recall the parameters ranges for the success of the NH approach in both weak- and strong-coupling regimes.

Within our approach we derived the Andreev NH Hamiltonian from the singular part of the barrier GF that depends also on the weight matrix, accounting for the quasiparticles weights in standard Landau theories. 
This matrix does not contribute to the Josephson current in the closed JJ limit. The result has been proved here to hold also for an open JJ, so that only the quasi-energies are required to describe the current.  
Considering that the NH Hamiltonian and the weight matrix do not commute in general, it would be interesting to consider their interplay within an analysis of the the quasi-ABS dynamics. 

Finally, we observe that this technique combines the advantage of an NH formalism with that of working with a reduced (barrier) GF, enabling the generalization of our approach to large-size or more complex barriers.
In particular, given the heuristics developed through our analysis, we envision its usage in systems comprising magnetic fields and spin-orbit interactions \cite{Zazunov2009, Yokoyama2014, Campagnano_2015, Minutillo2018, Maiellaro2024, Guarcello2024} and in interacting systems \cite{Rozhkov1999, Sellier2005, Cuevas2001}.
The approximation is expected to work for spectrally isolated levels and does not require the computation of the exact GF function but only of its singular part, whose computation might prove easier in some situations.

\begin{acknowledgments}
This work was supported by PNRR MUR project~PE0000023 - NQSTI, by the European Union's Horizon 2020 research and innovation programme under Grant Agreement No~101017733, by the MUR project~CN\_00000013-ICSC, by the  QuantERA II Programme STAQS project that has received funding from the European Union's Horizon 2020 research and innovation programme under Grant Agreement No~101017733, and by PRIN MUR Project TANQU 2022FLSPAJ. R.C. is supported by the European Union's Horizon EIC 2022 Pathfinder Challenges project “FERROMON-Ferrotransmons and Ferro-gatemons for Scalable Superconducting Quantum Computers” under Grant Agreement No~101115548.

The authors acknowledge C.W.J. Beenakker, R. Citro, R. Fazio, D. Giuliano and A. Russomanno for fruitful discussions and feedbacks.
\end{acknowledgments}

	\appendix

    \section{Green's function of the normal reservoir}
    \label{app: GFs_of_the_leads}    
\begin{figure}[htbp]
		\centering
		\includegraphics[scale=0.18]{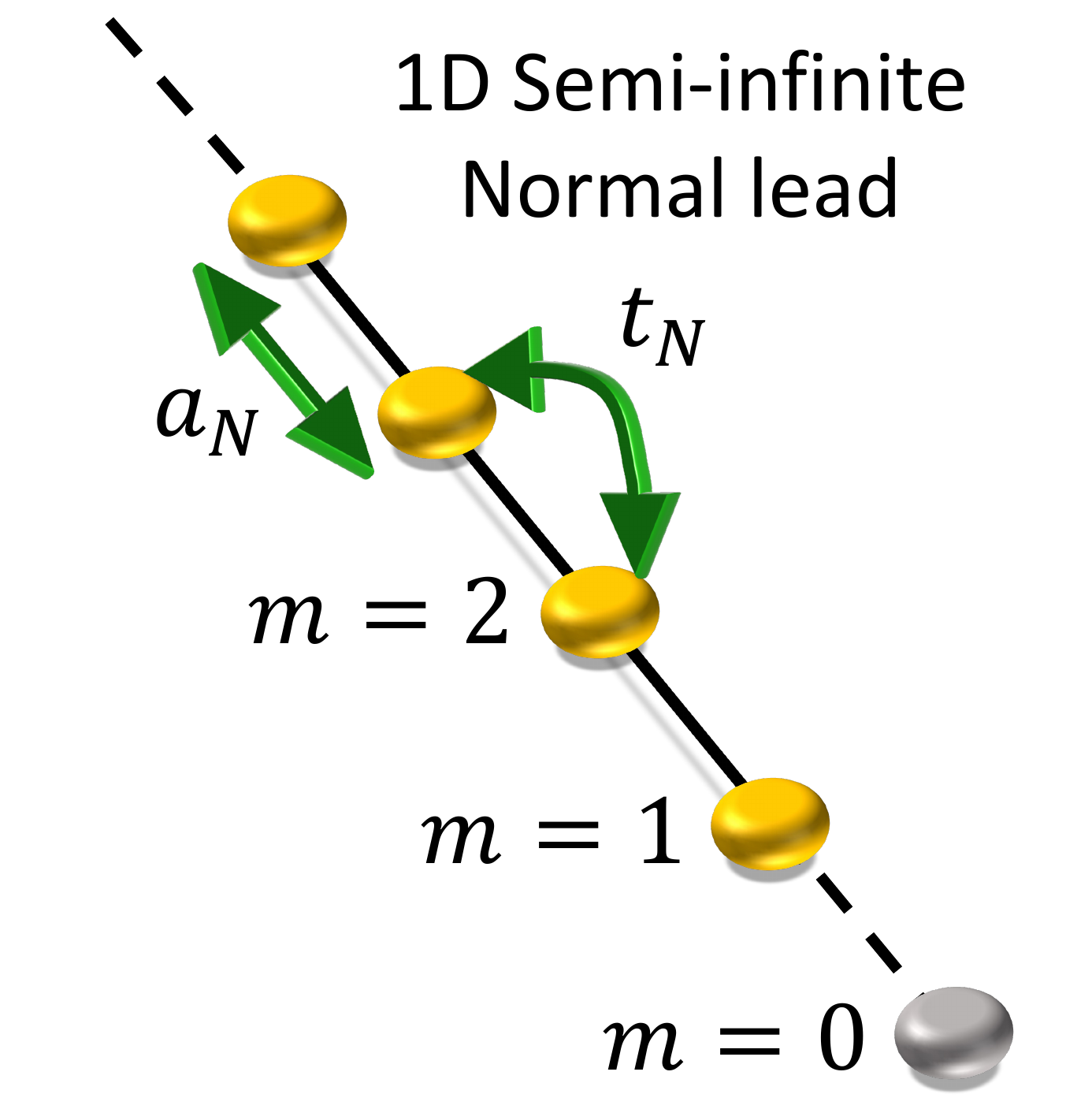}
		\caption{Sketch of the 1D semi-infinite normal lead. $t_{N}$ and $a_{N}$ are the hopping amplitude and the lattice constant, respectively. The hard wall boundary conditions imply the wave-function vanishing on the site $m=0$ in gray. }
		\label{1D_chain}
	\end{figure}

    	In this Appendix, we compute the both the Matsubara and retarded Green's function of the normal 1D semi-infinite lead in tight-binding formalism following Ref.~\cite{Ferry2009:book}. 
	We recall the N lead tight-binding Hamiltonian in Eq.~\ref{H_N_lead},
	\begin{equation}
		\begin{split}
			H_{\rm N}&=\sum_{m}\sum_{\sigma=\uparrow,\downarrow}\mu_{N}c_{N,m,\sigma}^{\dagger}c_{N,m,\sigma} + t_{N}c_{N,m+1,\sigma}^{\dagger}c_{N,m,\sigma}+\mathrm{H.c.}\\&=\sum_{\vec{k}}\sum_{\sigma=\uparrow,\downarrow}(\mu_N-\varepsilon_{N,\,\vec{k},\sigma}) c_{N,\vec{k},\sigma}^{\dagger}c_{N,\vec{k},\sigma}\,,
		\end{split}
	\end{equation}
    with $\mu_{N}$ being the normal metal chemical potential and $\varepsilon_{N,\, \vec{k},\sigma}=-2 t_{N} \cos{\left(k a_{N}\right)}$ being the dispersion law in momentum space. $t_{N}$ and $a_{N}$ are the hopping parameter and lattice constant of the 1D chain, respectively.
    The hard-wall boundary conditions requiring that the electronic wave-function vanishes at the site at $m=0$ (grey site in Fig.~\ref{1D_chain}), imply $\psi_{k}\left(m\right)=\sqrt{\frac{2}{\pi}}\sin\left(k a_{N} m\right)$.

	The influence of the semi-infinite chain can be condensed in the surface GF of the N lead that is the GF at $m=1$. Hence, we have the following GF expression 

    \begin{equation}
        \begin{split}   &G_{N}^{R}\left(m=m'=1,y=i\omega_{N},\,\omega+i\eta^{+}\right)=\\&\frac{2}{\pi}\int_{0}^{\pi}d\left(ka_{N}\right)\frac{\psi_{k}^{*}\left(m'\right)\psi_{k}\left(m\right)}{y-\mu_{N}-2t_{N}\cos{\left(ka_{N}\right)}}
        \end{split}
	\end{equation}

    where $y=i\omega_{n}$ for the Matsubara GF, $G_{N}(m=m'=1,i\omega_{n})$, and $y=\omega+i\eta^{+}$ for the retarded GF $G_{N}^{R}(m=m'=1,\omega)$,
	with $\eta^{+}$ being the usual convergence factor preserving causality. 
	These integrals have the form 
	\begin{equation}
		\frac{2}{\pi}\int_{0}^{\pi}\frac{\sin^{2}{\left(x\right)}}{a+b\cos{\left(x\right)}}\,,
        \label{integral_form}
	\end{equation}
	where $a=\omega+i\eta^{+}-\mu_{N}$ and $a=i\omega_{n}-\mu_{N}$ in the case of the retarded and Matsubara GF, respectively and $b=-2t_{N}$.
	For $a\in \mathbb{R}$, as in the case of the retarded GF, $G^{R}_{N}$ can be simply evaluated

	\begin{equation}
		\label{Retarded_N_lead_GF}
	   G_{N}^{R}\left(\omega\right)=\frac{1}{t_{N}}\left(\frac{\left(\omega-\mu_{N}\right)}{2t_{N}}-i\sqrt{1-\frac{\left(\omega-\mu_{N}\right)^2}{4t_{N}^{2}}}\right)\,.
	\end{equation}

	When $a\in \mathbb{C}$, as for the Matsubara GF, Eq.~\ref{integral_form} evaluates to 
    \begin{equation}
	   G_{N}\left(i\omega_{n}\right)=\frac{\left(i\omega_{n}-\mu_{N}\right)}{2t^{2}_{N}}-i\frac{\mathrm{sign}(\omega_{n})}{t_{N}}\sqrt{1-\frac{\left(i\omega_{n}-\mu_{N}\right)^2}{4t_{N}^{2}}}\,.
	\end{equation}
    This generalizes to the Nambu space as in the main text and in Ref.~\cite{JonckhereeMartin2009}:
	\begin{equation}
        \hat{G}_{N}\left(i\omega_{n}\right)=\frac{\left(i\omega_{n}-\mu_{N}\hat{\tau}_{z}\right)}{2t_{N}^2}-\frac{i\mathrm{sign}\left(\omega_{n}\right)}{t_{N}}\sqrt{1-\frac{\left(i\omega_{n}-\mu_{N}\hat{\tau}_{z}\right)^2}{4t_{N}^{2}}}\,. 
	\end{equation}
	
		We notice that the retarded Green’s function in Eq.~\ref{Retarded_N_lead_GF} is equivalent to the Eq.~31 in Ref.~\cite{Ryndyk2009} where the self-energy accounting for the dot - N lead interaction reads
		$\Sigma_{N}^{R}(\omega)=H_{\mathrm{T_{N}}}G^{R}_{N}H_{\mathrm{T_{N}}}^{T}$.
		It is worth mentioning that its real and imaginary parts differ when considering frequencies inside or outside the N lead band-width, i.e. $\omega-\mu_{N}<2 t_{N}$ or $\omega-\mu_{N}>2 t_{N}$. 
		They respectively describe the energy shift of the dot level and its broadening and have the following form:
		\begin{equation}
			\begin{split}
				\Re\left(\Sigma_{N}^{R}(\omega)\right)=&\frac{\gamma_{N}^2}{t_{N}}\frac{\left(\omega-\mu_{N}\right)}{2t_{N}}-\frac{\gamma_{N}^2}{t_{N}}\sqrt{\left(\frac{\omega-\mu_{N}}{2t_{N}}\right)^2-1}\times \\&
				\left[\Theta\left(\frac{\omega-\mu_{N}}{2t_{N}}-1\right)- \Theta\left(\frac{\mu_{N}-\omega}{2t_{N}}-1\right)\right],\\
			\end{split}
		\end{equation}
		\begin{equation}
			\Im\left(\Sigma_{N}^{R}(\omega)\right)=-\frac{\gamma_{N}^2}{t_{N}}\sqrt{1-\left(\frac{\omega-\mu_{N}}{2t_{N}}\right)^2}\Theta\left(1 - \left\vert\frac{\omega-\mu_{N}}{2t_{N}}\right\vert\right) \,.
		\end{equation}
		We note that for $\omega-\mu_{N}\ll2 t_{N}$, i.e. in the broad-band limit, there is no energy shift of the dot level and its broadening is frequency independent, while for  $\omega-\mu_{N}>2 t_{N}$ we have $\Im\left(\Sigma_{N}^{R}\right)=0$ such that no broadening is induced on the dot level.
    
    \section{Current formula on each lead for the T-junction}
    \label{Current in the SQDNS JJ}
	The current flowing through a system composed by three leads and a quantum dot, as in Fig.~\ref{fig: 1_SQDNS}, can be derived by the means of the tunneling Hamiltonian method~\cite{Zagoskin, Datta_1995, Cuevas1996}. 
	In the case of Josephson junctions, the current is, in principle, driven by both the voltage bias $V$ between the S and N leads and the superconducting phase difference $\phi=\phi_{R}-\phi_{L}$ between the superconductors.

    In equilibrium conditions, we assume $V=0$ between all the couples of leads, or simply $\mu_{L}=\mu_{R}=\mu_{N}=\mu=0$, but the calculation can be generalized to the finite voltage case~\cite{Zagoskin}.

    The procedure to compute the current flowing from the dot to the lead $i=L,R,N$ is the same for S and N leads~\cite{Zagoskin, Sellier2005}. 

	We start from the case of a normal lead, then we extend the current formula to case of a superconducting electrode by extending the dot and leads Green's functions to the Nambu space.
    
	In the case no bias voltage is applied to the banks, the current flowing from the dot to the lead $i$ in the Heisenberg picture reads as~\cite{Zagoskin}

	\begin{equation}
		J_{d,i}\left(s\right)=-|e|\left\langle\frac{d N_{i}(s)}{ds}\right\rangle=-ei\langle\left[H,N_{i}(s)\right]\rangle
		\label{Current_Def}
	\end{equation}
	where $e$ is the electron charge, $N_{i}(s)=\sum_{\vec{k},\sigma} c^{\dagger}_{i, \vec{k}, \sigma}(s)c_{i, \vec{k},\sigma}(s)$ is the electron number operator,  in the state $\vec{k}$ and spin $\sigma$, on the $i$ electrode at time $s$.  In the hypothesis of time-translational invariance the equal time correlations $\langle c^{\dagger}_{i, \vec{k},\sigma}(s)d_{\sigma}(s)\rangle$ and $\langle d^{\dagger}_{\sigma}(s)c_{i, \vec{k},\sigma}(s) \rangle$ can be computed at time $s=0$. 
    For the sake of simplicity, we take the flat-band approximation for the electrodes~\cite{Meng2009_PRB}. In this way, we can omit the sum over the quasi-momentum states $\vec{k}$ in $N_{i}(s)$ and define the tunneling Hamiltonian that couples the lead $i$ and the dot simply as $H_{T_{i}}=\gamma_{i}(c^{\dagger}_{i,\sigma}d_{\sigma}+\mathrm{H.c.})$, where $\gamma_{i}\in \mathbb{R}$. Hence, Eq.~\ref{Current_Def} becomes
   
	\begin{equation}
        \begin{split}
		J_{d,i}&=
		-i|e|\gamma_{i}\sum_{\sigma}\left(\left\langle c^{\dagger}_{i,\sigma}d_{\sigma}\right\rangle -  \left\langle d^{\dagger}_{\sigma}c_{i,\sigma} \right\rangle\right)\\
        &=-i|e|\gamma_{i}\sum_{\sigma}\left( G^{<}_{c_{i},d}-G^{<}_{d,c_{i}}\right),
        \end{split}
		\label{Current_hopp_2}
	\end{equation}
	where the correlation functions
    $G^{<}_{c_{i},d}$ and $G^{<}_{d, c_{i}}$ are the so-called	\emph{lesser Green's functions} in Non-equilibrium Green's function formalism~\cite{Zagoskin,Sellier2005,JonckhereeMartin2009, Debnath2024}. 
    
    These correlators describing the propagation of electrons between the dot and the lead $i$, at equilibrium and zero voltage bias,
	coincide with the dot-lead and lead-dot Green's functions, $G_{c_{i},d}$, $G_{d,c_{i}}$, in Matsubara representation.
    
	In a junction with two S electrodes and one N lead, the current on each lead can be obtained by extending the leads-dot GF to the Nambu space
    ~\cite{Asano2019, Furusaki1994, Asano2001, Minutillo2021, Ando_1991, Ahmad2022}
		\begin{equation}
		\label{Current_Matsubara}
		J_{i,d}\left(\phi\right)= - \dfrac{i e}{2} T \sum_{\omega_{n}} \mathrm{Tr} \left[\hat{\tau}_{z} \left(\hat{H}_{T_{i}}\hat{G}_{c_{i},d} (i\omega_{n}) -  \hat{H}_{T_{i}}\hat{G}_{d,c_{i}} (i\omega_{n})\right)   \right] \; ,
	\end{equation}
	where $\mathrm{Tr}$ stands for the trace over the Nambu space, $T$ is the temperature and $\omega_{n}=(2n+1)\pi T$ are the fermionic Matsubara frequencies.
	Analogously to $\hat{G}_{d}$ in Eq.~\ref{Gd_Dyson_equation_explicit}, also the GFs connecting leads and dot, $\hat{G}_{c_{i}d}$/$\hat{G}_{dc_{i}}$, can be calculated by the means of a Dyson equation~\cite{Zagoskin, Furusaki1994, Asano2001, Asano2019, Minutillo2021, Ahmad2022, Capecelatro2023}. Their expression in terms of $\hat{G}_{i}^{0}$ and $\hat{G}_{d}$ are, respectively: 
	\begin{equation}
		\begin{split}
			&\hat{G}_{dc_{i}}= \hat{G}_{d}\hat{H}_{T_{i}}\hat{G}_{i}^{0},  \\
			&\hat{G}_{c_{i}d}= \hat{G}_{i}^{0} \hat{H}_{T_{i}}\hat{G}_{d}\, .
		\end{split}
		\label{Lead-Dot_GFs}
	\end{equation}

    We recall that the leads/dot GF matrices in the Nambu space have the following structure 
    \begin{equation}
        \label{Nambu_structure_GF}
        \hat{G}_{i/d}\left(i\omega_{n}\right)=
        \begin{pmatrix}
            G_{i/d} & F_{i/d} \\
            F_{i/d}^{*} & - G_{i/d}^{*}
        \end{pmatrix} \,,
    \end{equation}
    
    where the diagonal elements, i.e. $\left(\hat{G}_{i/d}\right)_{1,1}=G_{i/d}(i\omega_{n})$ and $\left(\hat{G}_{i/d}\right)_{2,2}=- G_{i/d}^{*}(i\omega_{n})$, represent the particle and holes correlation functions, respectively, while the off-diagonal elements, i.e. $\left(\hat{G}_{i/d}\right)_{1,2}= F_{i/d}(i\omega_{n})$ and $\left(\hat{G}_{i/d}\right)_{2,2}= F_{i/d}^{*}(i\omega_{n})$, account for the superconducting correlations on the leads/dot, and are called \emph{anomalous} GFs. 

    Exploiting Eqs.~\ref{H_hopping}, \ref{Lead-Dot_GFs} and \ref{Nambu_structure_GF}, the current flowing between the lead $i$ and the dot, i.e. Eq.~\ref{Current_Matsubara}, can be further simplified
	\begin{equation}
		\label{Current_Matsubara_Simp_chap_4}
		J_{i,d} = \dfrac{i e \gamma_{i}^{2}}{2} T \sum_{\omega_{n}} \left[F_{d} (i\omega_{n}) F_{i}^{0,*} (i\omega_{n}) -  F_{i}^{0} (i\omega_{n}) F_{d}^{*} (i\omega_{n})  \right] \; ,
	\end{equation}
	where the trace over the Nambu space has been performed. At equilibrium only the superconducting correlation functions of the dot and the bare lead $i$, $F_{d}$ and $F_{i}^{0}$, contribute to the charge transfer across the junction~\cite{Sellier2005}.

	It is now easy to show that the current  flowing from the dot to the normal lead, $J_{N,d}$, in equilibrium conditions simply vanishes
	\begin{equation}
		J_{N,d}=-\frac{ieT\gamma_{N}^{2}}{2}\sum_{\omega_{n}}\mathrm{Tr}\left[\hat{G}_{d}\hat{\tau}_{z}\hat{G}_{N}^{0}-\hat{G}_{N}^{0}\hat{\tau}_{z}\hat{G}_{d}\right]=0\,,
	\end{equation}
    since there are no superconducting correlations on the normal lead.
    The only effect of the N lead on the SQDS JJ is to induce decoherence on the system.
	From the Kirchoff law, we find that $J_{L,d}=-J_{R,d}$, and that the Josephson current driven by the superconducting phase difference $\phi=\phi_{R}-\phi_{L}$ is
    \begin{equation}
		\label{Current_Matsubara_FF}
		J(\phi) = \dfrac{i e \gamma^{2}}{2} T \sum_{\omega_{n}} \left[F_{d} (i\omega_{n}) F_{R}^{0,*} (i\omega_{n}) -  F_{R}^{0} (i\omega_{n}) F_{d}^{*} (i\omega_{n})  \right] \; .
	\end{equation}
	
	Finally, we can write the more clear-cut CPR in Eq.~\ref{Current_Matsubara_FF} by using the expressions for $ F_{d}$ and $F_{R}^{0}$, in Eqs.~\ref{Bare_lead_GF_chap_4} and \ref{Gmatsu}~\cite{Sellier2005, Benjamin2007, Meng2009_PRB,  Beenakker1992, Capecelatro2023}.

Eq.~\ref{Current_Matsubara_FF}, when performing the analytic continuation of the anomalous dot Green's function, for $i\omega_{n}\rightarrow\omega +i\eta^{+}$, yields the Eqs.\ref{Andreev levels current}-\ref{Quasiparticles current} allowing to compute the sub-gap and supra-gap contributions separately.

 At equilibrium, the CPR can also be computed by deriving with respect to $\phi$ the dot free energy, $\mathcal{F}$, as $J\left(\phi\right)=2e\partial_{\phi}\mathcal{F}$.
 $\mathcal{F}$ involves both discrete and continuum spectral contributions~\cite{Benjamin2007, Meng2009_PRB, Beenakker1992, JonckhereeMartin2009, Rozhkov1999} and reads 
	\begin{equation}
		\label{Free_energy}
		\mathcal{F}(\phi)=-T\sum_{\omega_{n}}\ln{\left(-\det\hat{G}_{d}^{(-1)}\left(i\omega_{n}\right)\right)}\,.
	\end{equation}

    This definition of free energy in Matsubara representation, for a closed junction, remains valid in the case of an open junction at equilibrium, since it represents the total energy stored in the junction as a function of the phase difference between the two S leads.
    
	\section{Analytic continuations of the retarded Green's function}
	
	\subsection{Continuations of the retarded Green's function}
	\label{app: contin}
	The analytic continuation of the Matsubara Green's function in Eq.~\ref{Gmatsu} can be rewritten as 
	
	\begin{widetext}
		\begin{equation}
			\label{P_matrix_Nambu_analCont}
			\hat{G}_{d}(z)=
			\begin{pmatrix}
				&z-\varepsilon_{d}+\Gamma\frac{z}{s(\Delta,z)}+\frac{\Gamma_{N}}{2 t_{N}}z\left(\frac{s(2t_N,z)}{s(0,z)}-1\right) & \frac{\Gamma\Delta \cos{\left(\phi/2\right)}}{s(\Delta,z)} \\
				&\frac{\Gamma\Delta \cos{\left(\phi/2\right)}}{s(\Delta,z)}&
				z+\varepsilon_{d}+\Gamma\frac{z}{s(\Delta,z)}+\frac{\Gamma_{N}}{2 t_{N}}z\left(\frac{s(2t_N,z)}{s(0,z)}-1\right)
			\end{pmatrix}^{-1}
		\end{equation}
	\end{widetext}
	where $s(x,z) = \sqrt{x^2-z^2}$ and the sign function has been pre-processed as $\mathrm{sign}(\omega_{n}) = \omega_{n}/\sqrt{\omega_{n}^2} \rightarrow -iz/\sqrt{-z^2}$.
	
	Let us consider the limits in the definition of retarded and advanced Green's functions: $z\rightarrow \omega + i 0^\pm$. The function $s(x,z)$ has a branch point at $x$ and a branch cut running along the whole real axis and we have 
	\[
	s(x,\omega + i0^\pm) = \begin{cases}
		s(x,\omega) & |\omega| < x \\
		\tilde s(x,\omega) & |\omega| > x
	\end{cases}
	\]
	where we defined $\tilde s(x,\omega) = \mp i \omega \sqrt{\omega^2-x^2}/\sqrt{\omega^2}$.
	In the following we focus on the retarded function only, the advanced one having similar properties.
	Since in the Eq.~\ref{P_matrix_Nambu_analCont} three values for $x$ appear, either $0$, $\Delta$ and $t_N$, the retarded function has a different definition depending on whether $\omega$ sits on the left or on the right of the branch points (it always sits on the right of $0$).
		The piecewise definition of $G^R$ on the real axis is given in Eq.~\ref{GRpiecewise}. Each of these pieces have a different analytical continuation on the complex plane. So, allowing ourselves to write $\hat G_{d}(z) = \widetilde{ G}_{d}\left(s(0,z),s(\Delta,z),s(t_N,z)\right)$, the continuations of $G^R$ from each domain piece are
		\begin{eqnarray}
			\hat G^{R,in}(z) &=& \widetilde{ G}_{d}(\tilde s, s,s ) \nonumber\\
			\hat G^{R,out_1}(z) &=& \widetilde{ G}_{d}(\tilde s,\tilde s,s ) \nonumber\\
			\hat G^{R(out_2)}(z) &=& \widetilde{ G}_{d}(\tilde s,\tilde s,\tilde s )
		\end{eqnarray}
		where we have omitted the obvious dependences of the function $s$.
		
		For instance, the first row of the inverse of the sub-gap retarded Green's function reads as
		\begin{eqnarray}
  \label{GRspec}
			\left(\hat{G}^{R,in}(z)^{-1}\right)_{11}
			&=& z-\varepsilon_{d}+\Gamma\frac{z}{ \sqrt{\Delta^2-z^2}} + \nonumber\\
			&& +\frac{\Gamma_{N}}{2 t_{N}}\left( i\sqrt{4 t_N^2-z^2}-z\right) \nonumber\\
			\left(\hat{G}^{R,in}(z)^{-1}\right)_{12} &=& \frac{\Gamma\Delta \cos{\left(\phi/2\right)}}{\sqrt{\Delta^2-z^2}}, 
		\end{eqnarray}
		whereas the supra-gap retarded Green's function reads as
		\begin{eqnarray}
			\left(\hat{G}^{R,out_1}(z)^{-1}\right)_{11}
			&=& z-\varepsilon_{d}+i\Gamma\frac{\sqrt{z^2}}{\sqrt{z^2-\Delta^2}} + \nonumber\\
			&& +\frac{\Gamma_{N}}{2 t_{N}}\left( i\sqrt{4 t_N^2-z^2}-z\right) \nonumber\\
			\left(\hat{G}^{R,out_1}(z)^{-1}\right)_{12} &=& \frac{i \sqrt{z^2}\,\Gamma\Delta \cos{\left(\phi/2\right)}}{z\sqrt{z^2-\Delta^2}} 
		\end{eqnarray}
		
\begin{figure*} [ht!]
			\centering
			\includegraphics[width=.9\linewidth]{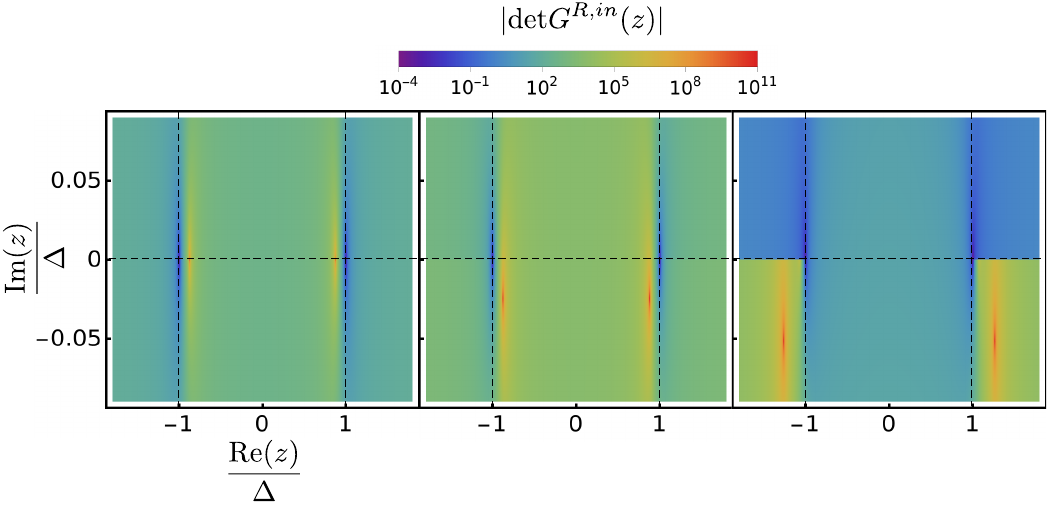} \\
			\caption{Analytic structure of the sub-gap retarded Green's functions from the matrix determinant for different regimes of the hybrization with the normal lead in the strong-coupling regime. (left)  Closed junction $\Gamma_N = 0$, (middle) NH system approximation regime $\Gamma_N = 0.05 \Gamma$ and (right) hybridized regime $\Gamma_N = 0.5 \Gamma$. The system parameters are $\varepsilon_d = 0,\, \Delta = 0.01,\, t_N = 10$ (in units of $\Gamma$) and $\phi = 1 \,\mathrm{rad}$. The poles move from sub-gap real frequencies to supra-gap values and acquire an imaginary part as the hybridization increase. The branch cuts are visible along supra-gap frequencies.}
			\label{fig:arg_abs_GR}
		\end{figure*}

		Notice that, at $\Gamma_N = 0$, $\hat{G}^{R,in} \equiv \hat{G}_{d}$ on the whole complex plane. In Fig.\ref{fig:arg_abs_GR} we show that $\hat{G}^{R,in}(z)$ features two poles and the branch cuts for real values of $z$ outside the gap.

       The retarded GFs $\hat{G}^{R(out_1/out_2)}$ account for the supra-gap continuum excitations.
       Thus, their branch-cut components are much more relevant in $\rho(\omega)$ and $J$, making the splitting into polar and branch-cut parts rather inconvenient. 
       This fact is evident even in the closed system: the supra-gap DOS profile cannot be approximated by any Lorentzian contribution, associated to the poles. Indeed, this part of the DOS is not related to interfacial properties of the junction but rather on the bulk specifics of superconducting leads. 
        
		\subsection{Continuum part of the retarded Green's function}
		\label{app: bc_funct}
  
		The continuum part of the retarded Green's function reads as \begin{equation}
        \label{Gbcut_formula}
			\hat G^{R,in}_{bcut}(z) = \frac{1}{2\pi i}\int_{|\omega|>\Delta} \frac{\Delta_{bcut} \hat G^{R,in}(\omega)}{\omega - z}
		\end{equation}
		where we defined the jump at the branch cut $\Delta_{bcut} \hat G^{R,in}(\omega) = \hat G^{R,in}(\omega+i 0^+) - \hat G^{R,in}(\omega-i 0^+)$. 
		Notice that the formula does not require the computation of the poles of the retarded function.
  
       In Fig.~\ref{fig: rho_bcut_check} (b) of the main text, we show that the expression above really complements the polar Green's function by computing their contributions to the DOS profile.
	   We justify the expression with the following argument. By taking the limits on the real axis, one can check that $\hat G^{R,in}_{bcut}$ has indeed a branch cut given by $\Delta_{bcut} \hat G^{R}$, the jump of $\hat G^{R}$. The difference between two analytic functions sharing only their branch-cuts locations and jump is necessarily analytic on the whole complex plane i.e. an entire function, by the Riemann's theorem on removable singularities (consider that the difference function is analytic right outside the branch-cut but is continuous across it). This entire function is the trivial function. Indeed the difference function must be bounded as it has no poles and other singularities would produce branch-cuts. However, the latter are removed by the difference procedure. Moreover, $\hat G^{R}$ vanishes at infinity (assuming some high energy cut-off is present), which implies that also the jump and $\hat G^{R,in}_{bcut}$, in turn, vanish. Thus, the entire function vanishes by Liouville's theorem.

       \subsection{Particle/hole symmetry}
       \label{app: symmetry}
		The system is symmetric under the particle-hole symmetry operator $K\,\hat{\tau}_y$ with $K$ the complex-conjugation operator. The symmetry implies 
\begin{eqnarray}
		\label{GR_props}
		\hat G^{R}(z) &=& -\hat{\tau}_y \,\hat G^{R}(- z^*)^* \,\hat{\tau}_y,
\end{eqnarray}
which applies individually to both polar and branch-cut components.
	
It is easy to see that the polar structure at the poles is fixed by the symmetry: if a pole is in $z=z_1$ with a residue $R_1$ and projector $\hat {P}_1$, then the other one is at $z_2=-z_1^*$ with $R_2 = R_1^*$ and $\hat {P}_2 = \hat{\tau}_y \hat {P}_1^* \hat{\tau}_y$. 
	
		\section{Determination of the effective non-Hermitian system } \label{app: eff_sys_details}
		In this appendix we collect results about the derivation of the NH Hamiltonian and its weight matrix and describe some of their properties by analyzing the Pauli-matrix structure of the retarded GF.
		
		\subsection{Derivation of the NH Hamiltonian and weight matrix}
        \label{app: DerivHZ}
		We can determine $\hat H_{eff}$ and $\hat Z$ by equating Eqs.~\ref{GR_eff} and \ref{GR_eff_neq}:
		\begin{eqnarray} 
			\label{compar}
        \hat Z\,(z - \hat H_{eff})^{-1}  &=& 
        \sum_{p=1,2} R_p \hat {P}_p(z - z_p)^{-1}.    
		\end{eqnarray}
		Looking at the big $|z|$ limit on both sides and retaining only $\mathcal{O}(|z|^{-1})$ terms, we recognize immediately that 
  \begin{equation}
  \label{Zeq}
  \hat Z= \sum_p R_p \hat {P}_p.
  \end{equation}
  
  To retrieve $H_{eff}$ we need to analyze the polar structure of the equality. The l.h.s. has poles at the eigenvalues of $H_{eff}$ which must equal the poles on the r.h.s. $z_p$. Taking the residues at one pole on both sides we obtain an equality between the residual matrices. In particular, we find $ \hat Z \hat P^H_p = R_{p}\hat P_{p}$ where $\hat P^H_p$ is the projector of the Hamiltonian associated to the eigenvalue $z_p$. Thus the effective NH Hamiltonian reads as 
 \begin{equation}
 \label{Heff_eq}
 \hat H_{eff} =\sum_p z_{ p} \,R_p \,\hat Z^{-1} \hat {P}_p.
 \end{equation}
  Observe that the Hamiltonian eigenvalues coincide with the pole of the retarded GF but the projectors of the former do not coincide necessarily with the projectors of the latter.

  Finally we comment on alternative definitions of the weight and the effective Hamitonian matrices. Clearly, we could have placed $Z$ on the right in Eq.~\ref{GR_eff_neq}. Otherwise we could have symmetrized the expression placing $\sqrt{Z}$ on the left and on the right~\cite{Bla23}, thereby introducing square-roots which, unfortunately, can be hard to handle. Yet another convention, used for instance in Ref.\cite{San24} is to take a Hadamard product of the weight matrix and the Green function of an isolated system, that is, posing $[\hat G^{R}_{pol}(z)]_{ab}= [(z - \widetilde{H}_{eff})^{-1}]_{ab} \,\widetilde{Z}_{ab}$. This definition is symmetric and quite appealing. However this representation leads to an inconsistency. Indeed, equating this expression with the Eq.\ref{GR_eff_neq} and integrating both sides along a complex contour encircling both $z_1, z_2$, it is possible to show that $\delta_{ab}\widetilde{Z}_{ab} = \hat Z_{ab}$ (with $\hat Z$ as per our definition). This would imply that the r.h.s. is diagonal which is not true in the open system, see the brief comment in App.\ref{app: ZandRes}.
  
   To delve deeper into the properties of the NH system, we must analyze the Pauli structure of the Green's function. This is the topic of the next paragraph.

 \subsection{Properties of $G^{R,in}_{pol}$}
 	\label{app: propGR}
 	We decompose the polar Green's function along the Pauli matrices basis as 
 	\begin{equation} \label{GF_pauli_decomp}
 		G^{R,in}_{pol}(z) = \xi(z) \hat 1 + \pmb v(z)\cdot \hat{\pmb \tau},
 	\end{equation}
 	where $\hat{\pmb \tau}=\left(\hat{\tau}_{x}, \hat{\tau}_{y}, \hat{\tau}_{z}\right)$ is the vector of Pauli matrices in Nambu space, $\xi(z)$ is a complex number and $\pmb v(z)$ a complex vector. Such can be further decomposed as $v(z) = \tilde\xi(z)(\pmb \alpha(z) + i \pmb\beta(z))$ with $\tilde\xi(z) = \sqrt{\pmb v(z) \cdot  \pmb v(z)}$ (no complex conjugation involved in the scalar product) and $\pmb \alpha(z)$ and $\pmb \beta(z)$ are real vectors satisfying $\alpha^2(z) - \beta^2(z) = 1$ and $\pmb \alpha(z) \cdot \pmb \beta(z) = 0$, with $\alpha(z)$ and $\beta(z)$ their norms. The decomposition is valid so long as $\tilde\xi(z) \neq 0$. 
 	When $\tilde\xi(z)$ vanishes the GF poles are degenerate. If, in addition, also the vector $\pmb v(z)$ vanishes then the degeneracy is of Hermitian type. Otherwise, it is easy to see that $z$ is an exceptional point and the decomposition of $G^{R,in}_{pol}$ in Eq.~\ref{GF_pauli_decomp} is no longer valid. 
 	In our system no exceptional point occurs and a degeneracy of hermitian type is allowed only at $\varepsilon_{d}=0$ and $\phi = \pi$, see next subsection. 
 	Importantly, since $G^R$ does not have $\hat \tau_y$ component, then also the $y$-components of the vectors $\pmb \alpha(z)$ and  $\pmb \beta(z)$ vanish. 
 	Notice that it holds $\pmb \alpha(z) \equiv (1,0,0)$ and $\pmb \beta(z) \equiv \pmb 0$ at $\varepsilon_d=0$. We relegate all comments on the $\Gamma_N = 0$ case in Section~\ref{app: NHredu}. 
 
 At each $z$ there are two complementary projectors onto the eigenvectors of the GF. They take the simple form
 	\[
 	\hat {P}_{\pm}(z) = \frac{1}{2} \left[ \hat 1 \pm (\pmb \alpha(z) + i \pmb\beta(z)) \cdot \hat{\pmb \tau} \right].
 	\]
 	It is possible to check that $G^{R,in}_{pol}(z) \hat {P}_{\pm}(z) = (\xi(z) \pm \tilde \xi(z)) \hat {P}_{\pm}(z)$.
 	We define $\hat {P}_{1}$ the projector associated to the pole $z_1$ and we assume it corresponds to the minus-sign projector $\hat {P}_{-}(z_1)$ (all considerations made here are independent from this assumption, when $\hat {P}_{1}$ is associated to the plus-sign projector all calculations are analogous) so that we have $\hat {P}_{1} = \frac{1}{2} \left[ \hat 1 - (\pmb \alpha(z) + i \pmb\beta(z)) \cdot \hat{\pmb \tau} \right]$ 
 	where hereafter we define $\pmb \alpha=\pmb \alpha(z_1)$ and $\pmb\beta = \pmb\beta(z_1)$. The projector 
 	$\hat {P}_{+}(z_1)$ complements $\hat P_{1}$ and is associated to the nonsigular eigenvalue of the GF. Defining $\hat {P}_{2}$ as the projector associated to the singular eigenvalue of the GF at $z = z_2$, we observe that particle/hole symmetry (see App.~\ref{app: symmetry}) implies $\hat {P}_{2} = \hat{\tau}{y} \hat {P}_{1}^* \hat \tau_y = \frac{1}{2} \left[ \hat 1 + (\pmb \alpha - i \pmb\beta) \cdot \hat{\pmb \tau} \right]$. At $\varepsilon_d = 0$, it holds $\hat {P}_{2} = \hat {P}_{1+}$ since $\pmb\beta = 0$. However, it is possible to check numerically that $\pmb\beta$ becomes finite at $\Gamma_{N}\neq0$ and $\varepsilon_{d}\neq0$.  Thus, one can check that 
 	\begin{equation}
 		\label{Pnon_com}
 		[\hat P_1,\hat P_2] \propto \alpha \beta \hat \tau_y
 	\end{equation} i.e. the residual matrices at the poles commute only when $\varepsilon_d = 0$.

 \subsection{Absence of exceptional points for the effective Hamiltonian} 		\label{app: EP}
 	Here, we prove that no exceptional point (EP) for the effective Hamiltonian will be present in the junction parameters manifold. We start off from the numerical evidence that the GF admits only one pair of particle/hole symmetric poles. By this symmetry, the presence of an EP requires the coalescence of the Andreev levels on the imaginary axis $z_1= z_2 = -i\lambda$.
	This is for example the case of a quantum dot Josephson junction coupled to a ferromagnetic metal lead analyzed in Ref.~\cite{CayaoSato2024}. We note that an EP for the Hamiltonian should be such also for the GF matrix in Eq.~\ref{GF_pauli_decomp}, which should occur if and only if $\tilde\xi(-i\lambda) = 0$, as mentioned in the paragraph above. By comparison of this equation with Eq.~\ref{Gmatsu}, it is easy to get the parameters of the decomposition at the EP: in particular $\pmb \alpha(-i\lambda) \neq  \pmb 0 $, $\pmb \beta(-i\lambda) = \pmb 0$, $\tilde\xi^2 = \frac{\Gamma^2\Delta^2}{\Delta^2+\lambda^2} \cos{\left(\frac{\phi}{2}\right)}^2   + \varepsilon_{d}^2$. At $\varepsilon_{d} \neq 0$ the latter quantity is always real and positive and does not vanish by varying any of the parameters in it (phase, coupling, complex imaginary frequency and detuning). At $\varepsilon_{d} = 0$ the quantity can vanish at $\phi = \pi$ but the degeneracy is of hermitian type. Therefore, we conclude that the effective Hamiltonian has no exceptional point.

  		\subsection{Properties of $\hat Z$ and residues in the weak and strong-coupling regimes} 		\label{app: ZandRes}

  The weight matrix, according to its definition in Eq.~\ref{Zeq} and the expression for the projectors given above, reads as 
		\begin{equation}
  \label{Zalphabeta}
		\hat Z = R^r\hat 1 - i (R^i \pmb \alpha + R^r \pmb \beta ) \cdot \hat{\pmb \tau},
		\end{equation}
	
		where we decompose $R_1 = R^r + i R^i$ along its real and imaginary part and use the property $R_2 = R_1^*$. The eigenvalues of $Z$ are complex: $R^r(1 \pm i \sqrt{(\alpha ||R||/R^r)^2 - 1})$, where we checked numerically that the square root is positive and vanishes only when the system is closed. Hence, the full matrix is not positive definite. The determinant instead is always positive and equals $\alpha ||R||$. 

  The fact that the determinant is real stems directly from the particle/hole symmetry and is not related to the specifics of our QD barrier. To see this, consider again Eq.~\ref{Zeq}. With a bigger barrier or with external fields (e.g. magnetic field) the expression stays the same save that the sum would run over more poles. Particle/hole symmetry implies (see the how $R_p$ and $\hat P_p$ transform under the symmetry below Eq.~\ref{GR_props}) that $Z = \hat \tau_y Z^* \hat \tau_y$, whence $\det(Z)=\det(Z)^*$ is real.

  We understand from Eq.~\ref{Zalphabeta} that $\hat Z$ is non-diagonal as soon as we have some coupling with the N lead. Indeed, at $\varepsilon_d = 0$, $\pmb \beta=0$ and $\pmb \alpha$ has a non-vanishing $x$-component whereas, at $\varepsilon_d \neq 0$, the two vectors are not parallel in general. This fact implies that the alternative definition for the weight and effective Hamiltonian matrices, $\widetilde{Z}$ and $\widetilde{H}_{eff}$, involving the Hadamard product is flawed in our case (see Appendix \ref{app: DerivHZ}).

\begin{figure} [ht!]
			\centering
			\includegraphics[width=.9\linewidth]{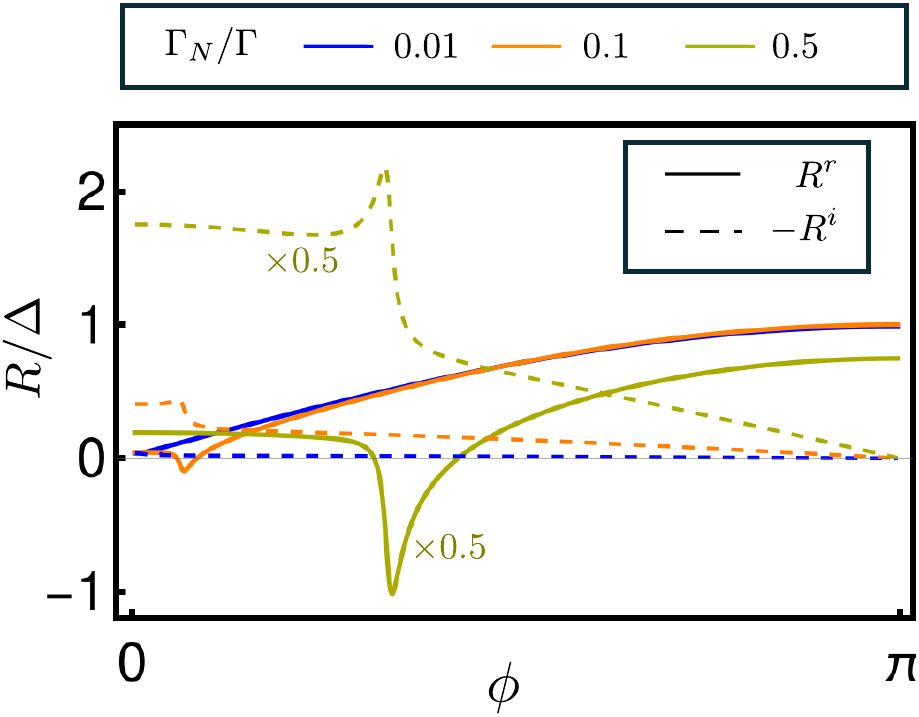} \\
			\caption{Real part, $R^r$, and imaginary part, $R^i$, of the residues in the strong-coupling regime for different ratios $\Gamma_N/\Gamma$. The curves at $\Gamma_N/\Gamma = 0.5$ have been halved. The system parameters are as in Fig.~\ref{fig: J_compar}}.
			\label{fig: residues}
		\end{figure}

In the weak-coupling regime, one can check that the weight matrix $\hat Z$ is close to the identity matrix. Instead, in the strong-coupling one, is much smaller (in matrix norm) as most part of the ABS is localized in the S leads. The imaginary part of the residues gets comparable to the real part as $\Gamma_N$ increases, determining a strong asymmetry in the DOS profile, see Fig.~\ref{fig: residues}). Notice in the same figure that the real part of the residues can get negative.

\subsection{From non-Hermiticity to Hermiticity at decoupled N lead}
\label{app: NHredu}

At $\Gamma_N = 0$ the retarded function $G^{R,in}(z)$ is real when $z$ lies on the real axis (cf. with Eq.~\ref{GRspec}). Thus, since its poles are located on the real axis, each residual matrix $R_p \hat P_p$ must be real as well. Upon tracing, we see that individually the residues and the projectors are real. Real projectors imply $\pmb \beta = 0$ and, in turn, $\hat {P}_{2} =\hat {P}_{1+}$ i.e. ${P}_{2}$ complements ${P}_{1}$ (see Appendix~\ref{app: propGR}). $\hat Z$ is real and proportional to the identity matrix by Eq.~\ref{Zalphabeta}. Finally, using Eq.~\ref{Heff_eq}, it is possible to see that the effective Hamiltonian is real and hermitian. 

  \subsection{Non-commutation of $\hat Z$ and $\hat H_{eff}$ at $\varepsilon_d\neq 0$} 
  \label{sec: non_comm}

The non-commutation of the weight matrix and the Hamiltonian at $\varepsilon_d\neq 0$ can be check directly using Eqs.~\ref{Zeq} and \ref{Heff_eq}. We have
\begin{eqnarray}
    [\hat Z,\hat H_{eff}] &=& \sum_{p} z_{p} \,R_p \,[\hat Z,\hat Z \hat {P}_p] \nonumber\\
    &=&  \hat Z \sum_p z_{ p} \,R_p \,[\hat Z, \hat {P}_p] \nonumber\\
    &=&  \hat Z \sum_{p,p'} z_{ p} \,R_p\,R_{p'} \,[\hat {P}_{p'}, \hat {P}_p] \nonumber\\
    &=&  \hat Z \sum_{p} z_{p} \,R_p \,R_{\bar p} \,[\hat {P}_{\bar p}, \hat {P}_p] \nonumber\\
    &=&  \hat Z \,|R_{1}|^2 \,[\hat {P}_{1}, \hat {P}_2](z_{2}-z_{1}) \nonumber\\
    &\propto&  \alpha \beta \,\hat Z \hat \tau_y \nonumber\\
\end{eqnarray}
 where $\bar p$ denotes the other index with respect to $p$; in the last step we used the result in Eq.~\ref{Pnon_com} and remove unimportant factors. It is clear that, since $\beta = 0$ only at $\varepsilon_d= 0$, the two matrices determining the NH system do not commute at finite dot detuning.

 \section{Further numerical results}
 \subsection{Josephson current in the non-resonant regime}
 \label{app: CPR_epsd_neq_0}
\begin{figure}
	\centering
    \includegraphics[scale=0.38]{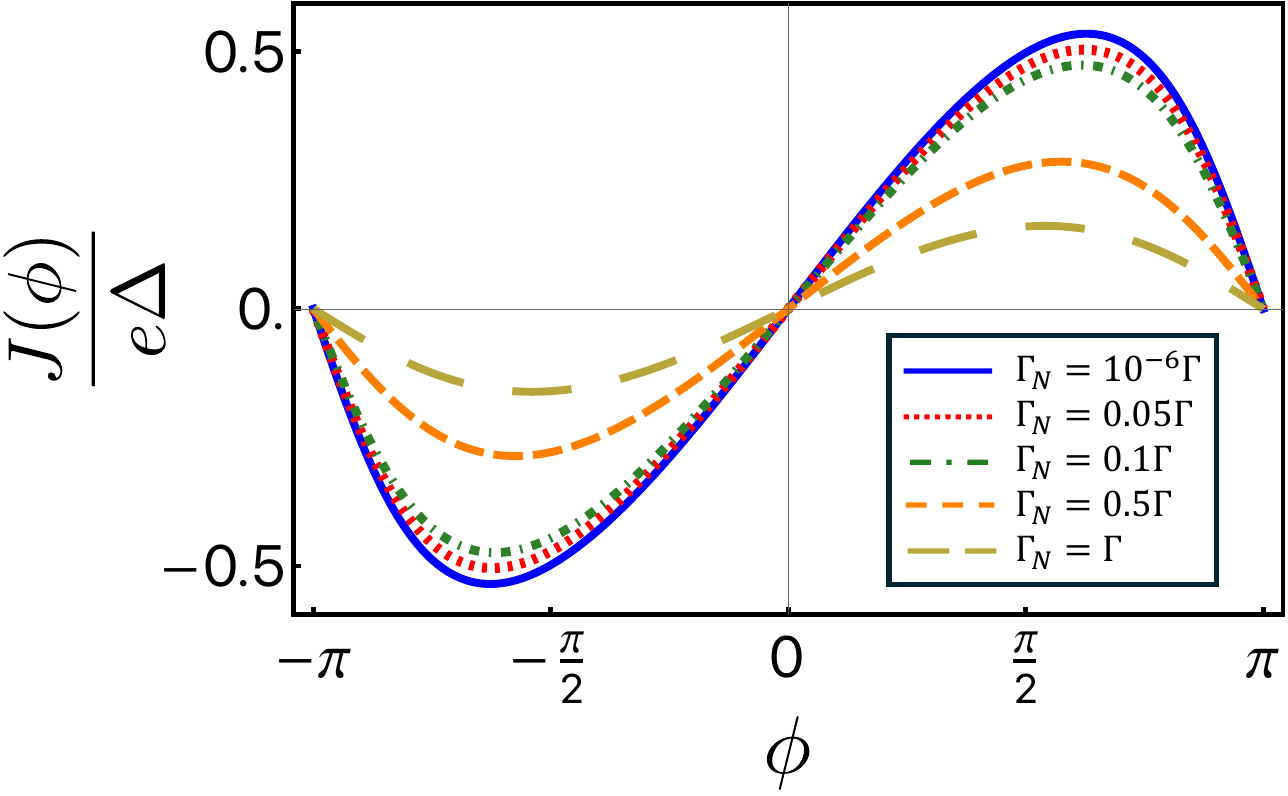}
	\caption{Current-phase relation (CPR) of the junction in the strong-coupling regime by varying the amplitude of the dot - N lead hybridization parameter $\Gamma_{N}$. The system parameters are (in units of $\Gamma$) $\varepsilon_{d}=0.5$, $\Delta=0.01$, $t_N=10$ and $T=10^{-3}\Delta$.}
	\label{Fig_CPR_offres_SC}
 \end{figure}

 Here we briefly comment how the Josephson current is affected by increasing the coupling with N in the case the quantum dot is tuned off resonance with respect to the superconducting leads.
 In Fig.~\ref{Fig_CPR_offres_SC}, we show the JJ CPR computed by the means of Eq.~\ref{Current_Matsubara_Final_chap_4} (in the strong-coupling regime) at $\varepsilon_{d}=0.5\,\Gamma$. This has a more sinusoidal shape as expected from the diminishing of the JJ transparency and the coupling with the N reservoir progressively lead to a lowering of the supercurrent as in the resonant tunneling.

 \label{Further_numerical_results}
    \subsection{Failure of the polar current formula in the weak-coupling regime}
    \label{Fail_Pol_weak_coupling}
    
    \begin{figure}[hbtp]
		\centering
		\includegraphics[scale=0.57]{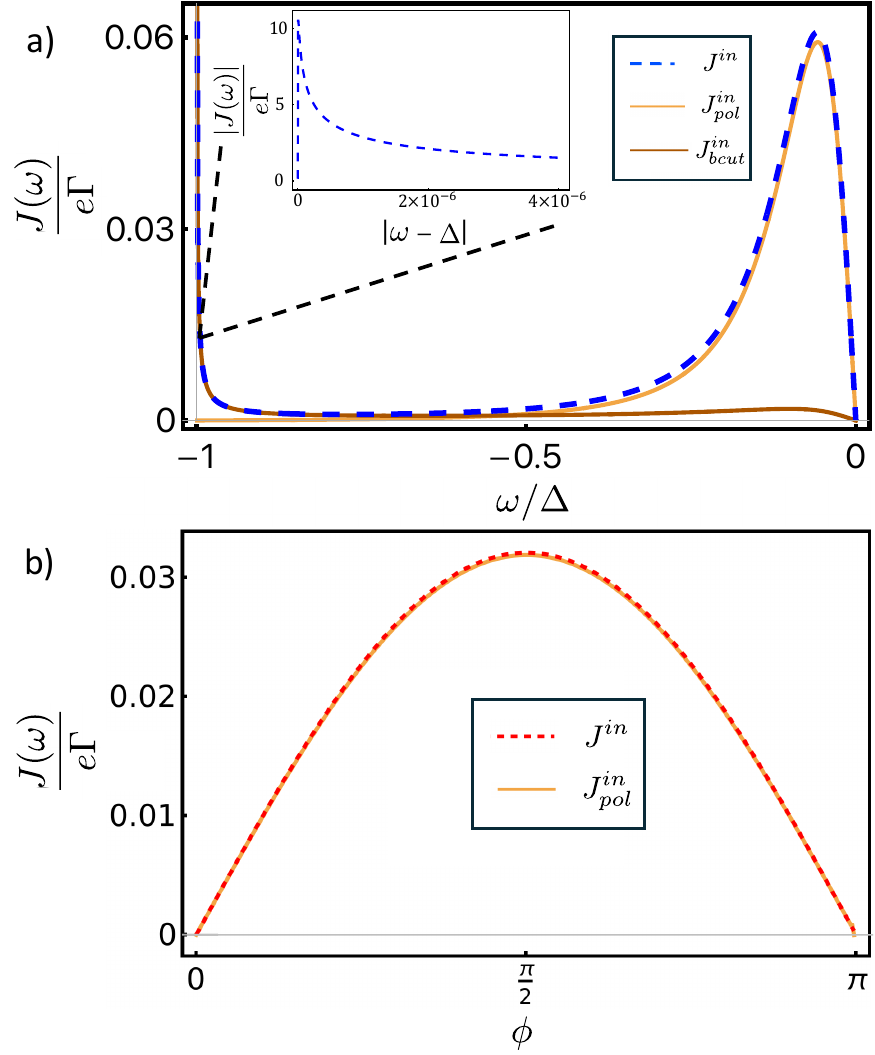}
		\caption{In the upper panel (a), we present the polar and branch-cut contributions to the current density $J(\omega)$ in the weak-coupling regime corresponding to the CPR curve in Fig.~\ref{fig: J_compar} computed at $\Gamma_{N}=10\,\Gamma$.
        The parameters of the junction are (in units of $\Delta$) $\epsilon_{d}=0$, $\Gamma=0.01$, $\phi=0.3$ and $T=0$.
        In the lower panel, we plot the exact CPR computed from the Eq.~\ref{Andreev levels current} by integrating over the reduced frequency defined by $\Delta_{ABS}=0.7\Delta$ and the polar current as in Eq.~\ref{Jpol_T0} finding a good agreement between the two curves. This result testifies the precision of the polar description in describing the current when the branch-cut contribution is excluded.} \label{fig:failure_weak_Jcomp}
    \end{figure}

	The polar current formulas in Eqs.~\ref{Jpol_general} and \ref{Jpol_T0} fail in predicting the exact sub-gap current in the weak-coupling regime as early as $\Gamma_{N}$ becomes a fractions of $\Delta$, Fig.~\ref{fig: J_compar} (first row, third plot from the right at $\Gamma_{N}=10\,\Gamma=0.1\,\Delta$).

    We analyze the mismatch by looking at the sub-gap current density $J^{in}(\omega)$ at a fixed phase in Fig.~\ref{fig:failure_weak_Jcomp} (a) to clarify the motivation of this lack of precision. The current density near the gap edge has a tiny contribution from the quasi-ABS and is almost entirely described by the contribution $J_{bcut}$ stemming from the reminder of the polar free energy $\mathcal{F}^{in} - \mathcal{F}^{in}_{pol}$.

    Therefore such density stems from the broadening of the supra-gap continuum contribution over the sub-gap frequency interval. The magnitude of this contribution increases linearly with $\Gamma_N$ (cf. the insets in Fig.~\ref{fig: 4_Jom_panel} and Fig.~\ref{fig:failure_weak_Jcomp}) in such a way that the phase window in which it is sizable is enlarged.
    As a consequence, by the same logic by which we analyzed the performance of the NH approximation only in the sub-gap region, it is reasonable to exclude those near-gap frequencies at which we see the non-ABS contribution. Thus, we define a \emph{effective} gap edge $\Delta_{ABS}\leq \Delta$ below which we integrate $J(\omega)$ in Eq.~\ref{Andreev levels current} and $J_{pol}(\omega)$ in Eq.~\ref{Jpol_general} to get a better match between the exact CPR and the approximated one.
    We apply this correction in Fig.~\ref{fig:failure_weak_Jcomp} (b) where the comparison among CPRs is made with $\Delta_{ABS}=0.7\Delta$. We observed a much better agreement between the two curves as expected. 
    This analysis of the current density confirms that the polar current can only describe Lorentzian-like resonances and performs well only if either the width of the branch-cut peak at $-\Delta$ is small or the sub-gap integration window is reduced.
    
    \subsection{NH approximation in the non-resonant regime}
    \label{app: epsd neq}
    \begin{figure}[t]
		\centering
        \includegraphics[scale=0.5]{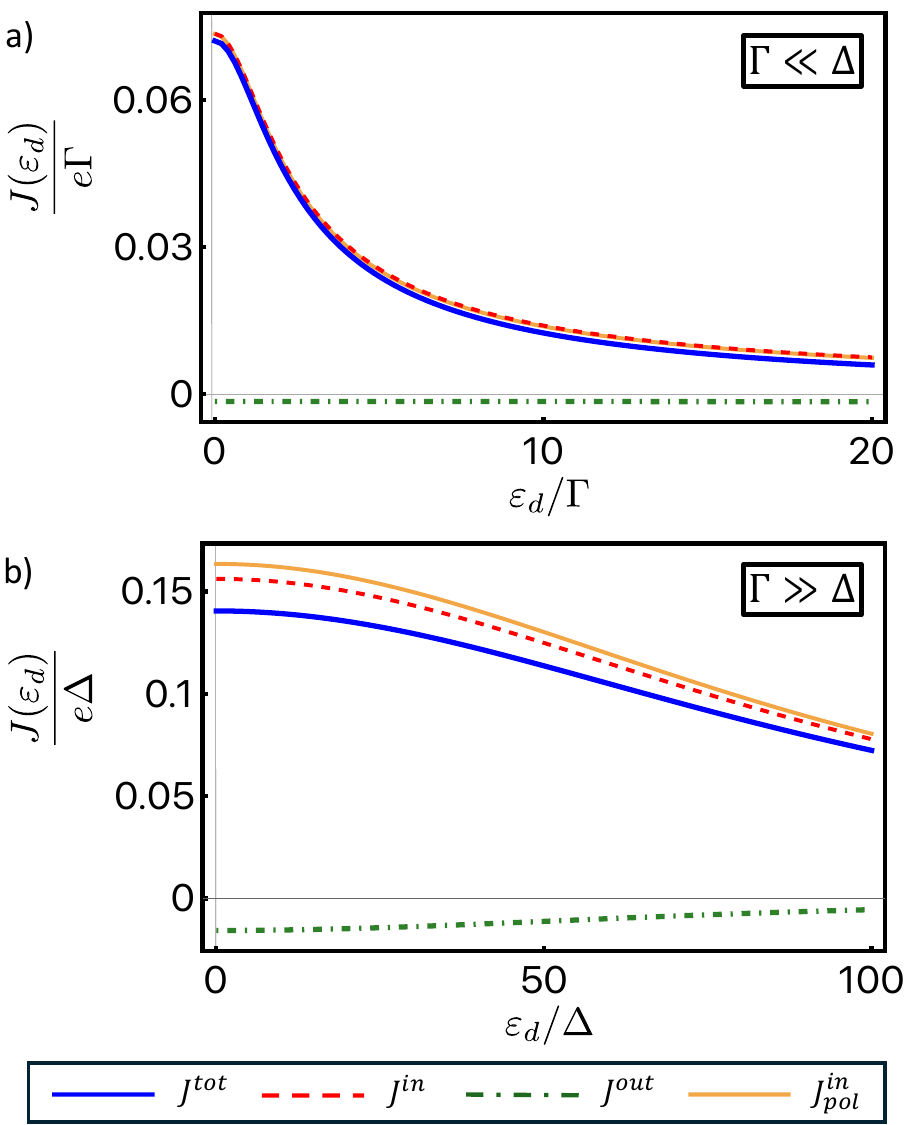}
        \caption{Performance of the NH approximation with finite detuning in the weak coupling (a) and strong coupling (b). The current is decomposed as in Fig.~\ref{fig: 3_Jin_Jout_panel} with the addition of the $J_{pol}^{in}$ (orange line) computed as in Eq.~\ref{Jpol_general} at $\phi=0.3$ and $T=0$. The parameters of the junction  are, respectively, $\varepsilon_{d}=0$, $t_{N}=10$,
        $\Gamma=0.01$,
        $\Gamma_{N}=0.01=\Gamma$; 
         (in units $\Delta$)
         for the weak coupling (a) and $\epsilon_{d}=0$,
         $t_{N}=10$, $\Delta=0.01$, $\Gamma_{N}=0.1=0.1\,\Gamma$ (in units $\Gamma$) for the strong coupling (b).}
        \label{fig: JvsEpsd}
    \end{figure}

    We show here that the heuristics about the applicability of the NH approximation developed for the resonant regime $\varepsilon_d=0$ applies also in the non-resonant one with $\varepsilon_d\neq0$.
    In Fig.~\ref{fig: JvsEpsd} we show the current contributions at fixed phase and coupling with the N lead, varying the dot local energy. 
    
    First of all, we notice that the effect of the detuning is analogous to that of decreasing the barrier transparency. In the weak coupling, see Fig.~\ref{fig: JvsEpsd}(a), $J^{out}(\varepsilon_d)$ is roughly constant while $J^{in}(\varepsilon_d)$ decreases sensibly (cf. Fig.~\ref{fig: 3_Jin_Jout_panel}(a)). Instead, in the strong coupling, see Fig.~\ref{fig: JvsEpsd}(b), both the $J^{in}$ and $J^{out}$ decrease with the dot detuning, similarly to what happens in the resonant case by increasing the coupling with N, (cf. Fig.~\ref{fig: 3_Jin_Jout_panel}(b)). 

    Concerning the performance of the NH approximation, we observe that the detuning does not disrupt it, instead it has either a neutral or beneficial effect. In the weak-coupling regime we choose a coupling $\Gamma_N$ at which the approximation is excellent at $\varepsilon_d =0$ (cf. with Fig.~\ref{fig: J_compar}(b)). We observe that the current from Eq.~\ref{Jpol_T0} coincides always with the exact one, as we expected, since the quasi-ABS sit always very distant from the gap edges. 
    In the strong coupling we set the parameters $\Gamma_N$ and $\phi$ to values at which the approximation is not good at $\varepsilon_d =0$ as the quasi-ABS sit close to the gap edges (cf. with Fig.~\ref{fig: J_compar}(e)). 
    
    Surprisingly, even though a finite $\varepsilon_d$ pushes the states toward the edges, the approximation improves. To understand this remarkable phenomenon we have to look at the imaginary part of the pole $\lambda$ of the retarded GF, which sets the size of the quasi-ABS Fano resonances. Increasing $\varepsilon_d$, $\lambda$ diminishes at a faster rate than that at which the real part of one pole $\epsilon$ approaches one edge. Next to it, we checked that we the current density does not bear any sensible modification at the gap edge of the kind shown in the previous paragraph Section~\ref{Fail_Pol_weak_coupling} when increasing the detuning. Thus, the Fano resonance gets closer but much sharper and with little overlap with the continuum contribution as the dot goes off-resonance. 
    To conclude, this analysis confirms that the intuition about the applicability of the NH approximation is valid also in the non-resonant regime. So long as the quasi-ABS do not overlap with continuum states the approximation works.

	\end{document}